\documentclass[preprints,article,accept,moreauthors,pdftex]{Definitions/mdpi}
\firstpage{1}
\pubvolume{1}
\issuenum{1}
\articlenumber{0}
\pubyear{XXXX}
\copyrightyear{XXXX}
\datereceived{XX Month Year}
\dateaccepted{XX Month Year}
\datepublished{}
\hreflink{https://doi.org/} 
\pdfoutput=1
\Title{Quantum-Mechanical Modelling of Asymmetric Opinion Polarisation in Social Networks}
\TitleCitation{Quantum-Mechanical Modelling of Asymmetric Opinion Polarisation in Social Networks}

\definecolor{my_green}{rgb}{0.55, 0.71, 0.0}

\newcommand{\Hamiltonian}{{\hat H}}
\newcommand{\icol}[1]{
  \left[\begin{smallmatrix}#1\end{smallmatrix}\right]%
}

\usepackage{textcomp}
\usepackage{amsmath, amssymb}
\usepackage{graphicx}
\usepackage{mathtools}
\usepackage{bm}
\usepackage{color}
\usepackage{comment}

\Author{Ivan S.~Maksymov$^{1}$\orcidA{} and Ganna~Pogrebna$^{1,2,3}$\orcidB{}}

\AuthorNames{Ivan S.~Maksymov, Ganna Pogrebna}

\AuthorCitation{Maksymov, I.S.; Pogrebna, G.}

\address{%
$^{1}$ \quad Artificial Intelligence and Cyber Futures Institute, Charles Sturt University, Bathurst, NSW 2795, Australia\\
$^{2}$ \quad The Alan Turing Institute, British Library, London NW1 2DB, United Kingdom\\
$^{3}$ \quad The University of Sydney Business School, Darlington NSW 2006, Australia}

\corres{Correspondence: imaksymov@csu.edu.au}

\abstract{We propose a quantum-mechanical model that represents a human system of beliefs as quantised energy levels of a physical system. This model underscores a novel perspective on opinion dynamics, recreating a broad range of experimental and real-world data that exhibit an asymmetry of opinion radicalisation. In particular, the model demonstrates the phenomena of pronounced conservatism versus mild liberalism when individuals are exposed to opposing views, mirroring recent findings on opinion polarisation via social media exposure. Advancing this model, we establish a solid framework that integrates elements from physics, psychology, behavioural science, decision-making theory and philosophy, and also emphasise the inherent advantages of the quantum approach over traditional classical models. We also suggest a number of new directions for future research work on quantum-mechanical models of human cognition and decision-making.}


\begin{document}
\section{Introduction}
The dynamics of human beliefs and radicalisation of opinions have emerged as defining factors in our society's trajectory, as witnessed across a myriad of political, religious, moral and racial divides \cite{ma2019psychological, dixon2019unintended, druckman2019evidence,van2021inoculating, horne2015countering, nyhan2015does, van2018partisan, Bai18, redlawsk2002hot, nyhan2010corrections, nyhan2021backfire, chen2021opinion, thomm2021preservice, cook2004firing, liebertz2021backfiring}. Evidence of this pervasive polarisation is abundant. For instance, we see political polarisation in the United States, where Democrats and Republicans hold increasingly disparate views on topics such as gun control, climate change, immigration and healthcare \cite{Bai18, hamilton2015trust}. Racial divides are also starkly apparent \cite{kinder2001exploring}, resulting in differing views on police brutality \cite{lawrence2022politics} and immigration policy \cite{hanson2005does}. In the context of religion, we observe polarisation in the form of divergent beliefs on topics such as abortion, LGBTQ+ rights and the role of religion in public life \cite{liebertz2021backfiring, Mak24_gender}. Ethical polarisation is also evident in ongoing debates on euthanasia, capital punishment and animal rights \cite{tatalovich2014moral}.

Yet, polarisation is not confined to these broad social issues since it extends into more specific realms such as differing views on vaccines, where pro-vaccine advocates and vaccine sceptics are at odds \cite{nyhan2015does, van2021inoculating}. Opinions on technological issues involving artificial intelligence, genetic engineering and data privacy are further examples \cite{abedin2022managing, urban2000attitudes}. We also observe polarisation in our responses to major global challenges, such as climate change \cite{dixon2019unintended, druckman2019evidence, hamilton2015trust}, where supporters and sceptics hold divergent views. Polarisation further manifests itself in societal attitudes towards economic disparities, income inequality and social safety nets \cite{montalvo2003religious}.

These divisions have spawned numerous problems ranging from unrest at the societal level to conflicts within smaller groups with opposing beliefs. People with conflicting beliefs are often regarded as adversaries, leading to divisive and destructive practices such as information warfare, cyber-attacks, derogatory comments in social media and the spreading of fake news \cite{vicario2019polarization}. While the damage inflicted to the information ecosystem by such actions is often unappreciated and underestimated, its impact on societal cohesion is profound and far-reaching.

This polarised landscape is significantly influenced by confirmation bias, a psychological phenomenon where individuals interpret or seek information that confirms their preexisting beliefs while overlooking contradictory evidence \cite{oswald2004confirmation}. Such a bias shapes the behaviour of social network users, leading them to select and propagate claims aligned with their beliefs, fuelling group polarisation and entrenchment of beliefs as well as reinforcing and perpetuating societal divides. Of particular interest is the backfire effect---a manifestation of confirmation bias, referring the tendency of people to give more credence to evidence that supports their preexisting beliefs \cite{Bai18}.

Early research into the backfire effect highlights that not only are fallacious beliefs stubbornly resistant to correction, but attempts to refute them can inadvertently cement them further \cite{nyhan2010corrections, nyhan2013hazards, nyhan2014effective, nyhan2015does, cameron2013patient, thorson2016belief, chan2017debunking}. This phenomenon extends to beliefs that align closely with an individual’s worldview; for instance, challenging principles tied to Republican ideologies might only deepen a Republican's adherence to those principles. More recent studies have suggested that the backfire effect may not be as universally consistent as earlier thought \cite{swire2020searching}. However, the current field consensus reaffirms that while refutations may temporarily sway beliefs toward accuracy, this impact is generally short-lived \cite{nyhan2021backfire}. The impermanence of these corrections means that inaccuracies, as well as beliefs reinforced by personal ideologies, can continue to influence public opinion well after being discredited. The resilience of such beliefs and misconceptions likely stems from the corrective information not being conveyed in a manner that results in a long-lasting change in perspective \cite{porter2023correcting}. This issue is compounded by the fact that individuals often navigate complex information through the lens of their preexisting worldviews and affiliations, which can override their interpretations and acceptance of new evidence\cite{ecker2022psychological}. 

To further understand the complex dynamics of the backfire effect, our work advances a model derived from the principles of quantum mechanics \cite{Mes62, Bus12}. This approach provides a unique perspective for comprehending and quantifying opinion radicalisation, elucidating the evolution of individual beliefs when exposed to counter-attitudinal information and revealing its contribution to societal polarisation.
\begin{figure}[t]
\centering
\includegraphics[width=.99\linewidth]{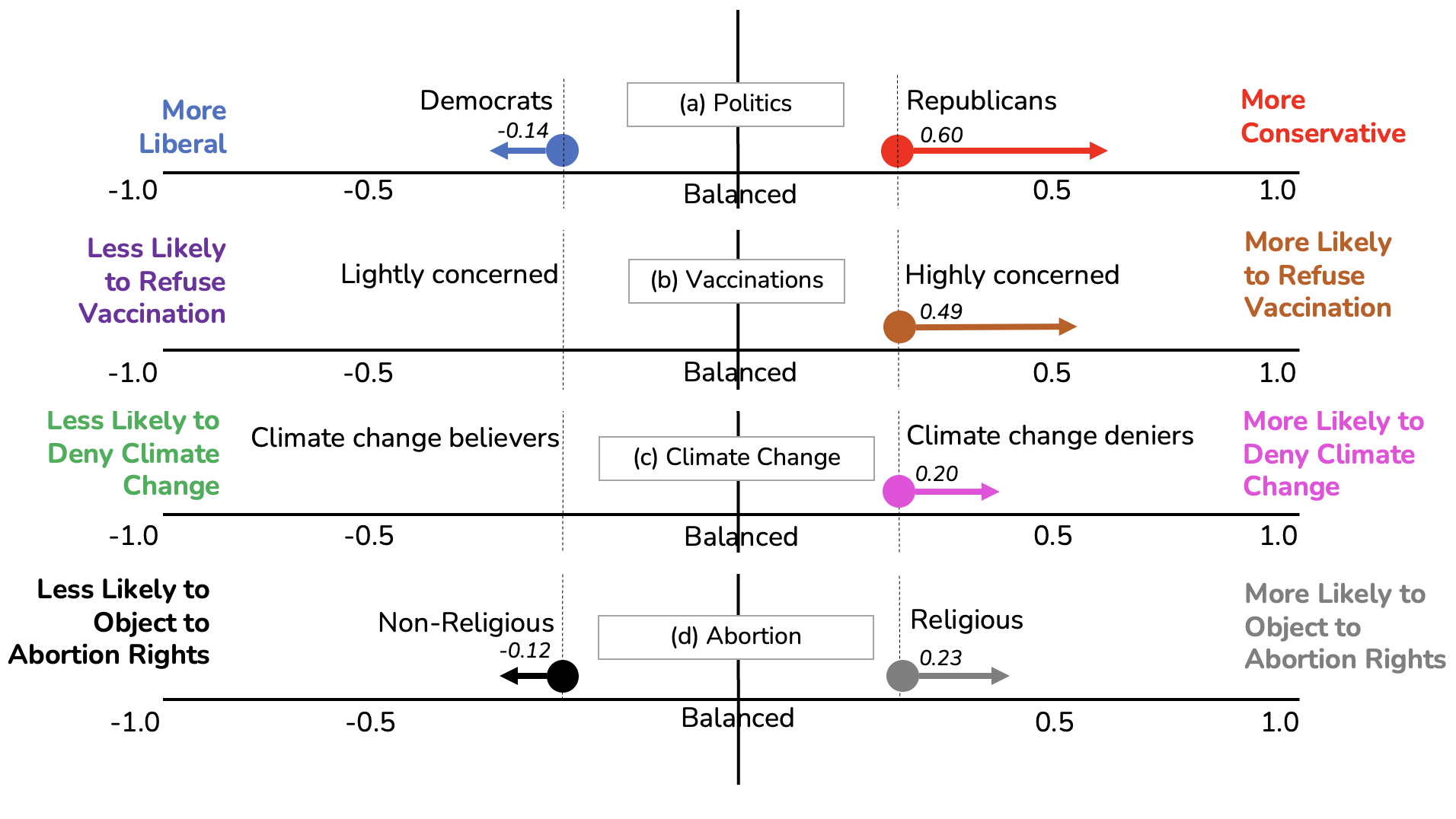}
\caption{Notable examples of the backfire effect in the literature. The figure represents the backfire effect observed in four studies covering {\bf(a)}~Politics \cite{Bai18}; {\bf(b)}~Vaccinations \cite{nyhan2015does}; {\bf(c)}~Climate Change \cite{dixon2019unintended}; and {\bf(d)}~Abortion \cite{liebertz2021backfiring}. In each of these studies, groups exhibited the backfire effect after being subjected to an opposing view. Note:~Each of the horizontal axes should be read independently. Results from the cited papers are normalised to be presented on a scale from -1~to~1. The figure demonstrates the direction and magnitude of the backfire effect for one or more groups in each study, showcasing that Republicans, people highly concerned with vaccination side effects, climate change deniers and religious people are more likely to become more conservative, more likely to refuse vaccination, more likely to deny climate change and more likely to object to the abortion rights, respectively, after being subjected to opposing views. At the same time, Democrats and non-religious people exhibit a much lower backfire effect in magnitude than their Republican and religious counterparts, respectively. The counterpart views are not available for all studies.}\label{fig:litexamples}
\end{figure}

Examining societal opinion dynamics through the lens of this quantum-inspired model bears considerable implications. This model reproduces the asymmetric behaviour of opinion radicalisation commonly observed when individuals confront opposing views in social media. Such a radicalisation, augmented by the backfire effect, is prominently manifested in political domains. For instance, in the political arena of the United States, exposure to counter-attitudinal information typically results in an entrenchment of beliefs among both Democrats and Republicans. However, Republicans have been observed to radicalise to a higher degree than Democrats in response to conflicting perspectives \cite{Bai18}.

It is noteworthy that the applications of the model are not limited to politics since the phenomenon of radicalisation applies to other societal issues such as climate change, gun control and vaccinations, often leading to an intensification of individuals' initial beliefs when they are presented with contrasting evidence or viewpoints. Our model elucidates this counter-intuitive dynamics, thus offering a new perspective on the complex landscape of human beliefs and opinions.

Figure~\ref{fig:litexamples} provides empirical evidence of the `double-sided' polarisation behaviour triggered by the backfire effect, observed in the domains of politics \cite{Bai18} and abortion \cite{liebertz2021backfiring}. Conversely, in Fig.~\ref{fig:litexamples} the domains of climate change \cite{dixon2019unintended} and vaccination \cite{nyhan2015does} are marked by a seemingly `one-sided' view. This discrepancy arises due to missing data for these areas as studies focusing on climate change and vaccinations typically present scientifically accurate information in experimental treatments without offering opposing viewpoints.

The apparent limitation of the available motivational data should not be perceived as a detriment to our investigation since our model can predict radicalisation trends in contexts where data on opposing views are unavailable or scarce. Thus, by elucidating such trends, we aspire to contribute to an interdisciplinary understanding of opinion dynamics, with the potential of informing interventions that could mitigate polarisation and foster more productive and respectful societal dialogues.

Furthermore, the proposed model reveals the fundamental origin of advantages over classical models in understanding the radicalisation of opinions, thereby providing a robust framework that can predict opinion changes with superior accuracy and reliability. This model also contributes to the growing body of research exploring the plausibility of the `quantum mind' hypothesis, which posits that the principles of quantum mechanics could underlie cognitive processes \cite{Khr06, Bus12, mindell2012quantum}. In particular, through its successful simulation of experimental data on opinion dynamics, the model provides further arguments supporting the potential of this hypothesis.

The examples of the backfire effect in Fig.~\ref{fig:litexamples} provide evidence of the scale and gravity of polarisation across multiple dimensions of our society. Yet, understanding these dynamics is only the first step. The ultimate goal is to leverage these insights and devise strategies to alleviate the destructive effects of radicalisation and polarisation. By offering a comprehensive and robust framework to interpret these dynamics, the proposed quantum-mechanical model stands to make a contribution to these efforts.

\section{Results}
\subsection{Quantised Energy Level Model of the System of Human Beliefs}
The application of the fundamental principles of quantum mechanics in the fields of behavioural science, economics and decision-making has opened up novel opportunities for unifying and formalising previously heuristically formulated psychological, cognitive and finance-related concepts and ideas \cite{Khr06, Bus12}. In particular, an analogy between mental states and quantum states has been used to explain cognitive dissonance \cite{Khr14} and gender fluidity \cite{Mak24_gender}. A similar approach has been adopted in models of confirmation bias \cite{All14, Gro17}, gambling \cite{Bus12, Zha17_1}, Prisoner's Dilemma game \cite{Bus12, Che06}, conjunction fallacy \cite{Bus12, Gro17} and Ellsberg paradox \cite{Bus12}. However, although there exist classical models of confirmation bias \cite{All14, Gro17}, opinion formation and polarisation \cite{Gal05, Cas09, Hu17, Eyr17, Del17, Red19, Bel20, Tok21, Cin21, Gal22, Hoh23, Int23, Geo23}, human morality \cite{Cap21} and backfire effect \cite{Che21, Axe21}, the core principles of quantum mechanics exploited in this paper have not been previously applied to the aforementioned psychological phenomena.

\begin{figure}[t]
\centering
\includegraphics[width=.75\linewidth]{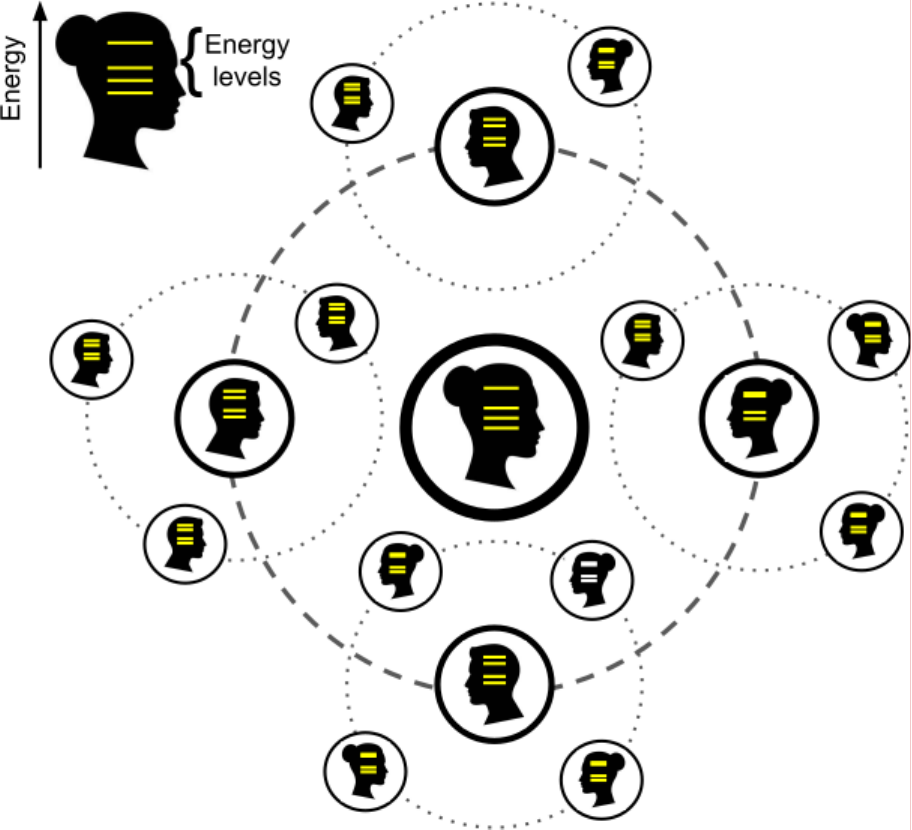}
\caption{Sketch of a social network showing the social circles and human systems of beliefs represented by discrete energy levels (the horizontal lines superposed on the human head silhouettes). The physical analogy between overlaps of social circles and overlaps of atomic orbitals and concomitant energy level changes underpins the model proposed in this paper.}
\label{fig:social_energy_levels}
\end{figure}

We propose a quantum-mechanical model of human system of beliefs (see the inset in Fig.~\ref{fig:social_energy_levels}) and demonstrate its application to opinion radicalisation and backfire effect in social networks. It is well-known that a quantum-mechanical system can have only certain energies levels compared with a continuum of energy states of a classical system \cite{Mes62}. Therefore, drawing on the previous works in the domain of socio-physics \cite{Aer22, Aer22_1}, we represent the system of human beliefs as a set of discrete energy levels. Furthermore, in Fig.~\ref{fig:social_energy_levels} we draw a hypothetical parallel between the social network circles and atomic orbitals used in quantum mechanics to describe the location and wave properties of an electron in an atom \cite{Kittel}. The electrons occupy atomic orbitals that have discrete energy levels \cite{Kittel}. When atoms interact, their atomic orbitals overlap and energy levels hybridise \cite{Kittel}. Extending the previous works that employ spin-like classical magnetic dipoles to understand and predict the social interaction between individuals \cite{Axe97, Szn00, Cas09, Red19, Mak24_gender}, we rigorously adopt the quantum-mechanical processes of atomic orbitals overlap and energy level hybridisation to model the influence of social network circles on the system of beliefs of an individual. 

The proposed model stands on solid psychological, philosophical, mathematical and physical ground. Firstly, it is consistent with both notion of discrete mental states \cite{Rou09, Khr18, Aer22, Lin23} and understanding of information as energy states of a physical system \cite{Toy10, Dit14, Vop19}. Secondly, the model captures psychological \cite{Uso16} and mathematical \cite{DeG74, Red19} complexity of a system of human beliefs. Thirdly, the model aligns with the famous philosophical maxim pronounced by Ortega y Gasset: 'Yo soy yo y mi circunstancia' (`I am me and my circumstance', see \cite{Ortega_Gasset}, p.~322; a link between Ortega y Gasset's circumstance and quantum aspects of cognition has also been established \cite{deC13}). In the model, the circumstance is represented by interactions between atoms and its influence results in a change of the energy level structure. Finally, the model relies on a rigorous solution of the fundamental Schr{\"o}dinger equation \cite{Mes62}, the relevance of which to the domains of psychology and decision-making was demonstrated \cite{Bus12, Mak24_illusions}.

\begin{figure}[t]
\centering
\includegraphics[width=.85\linewidth]{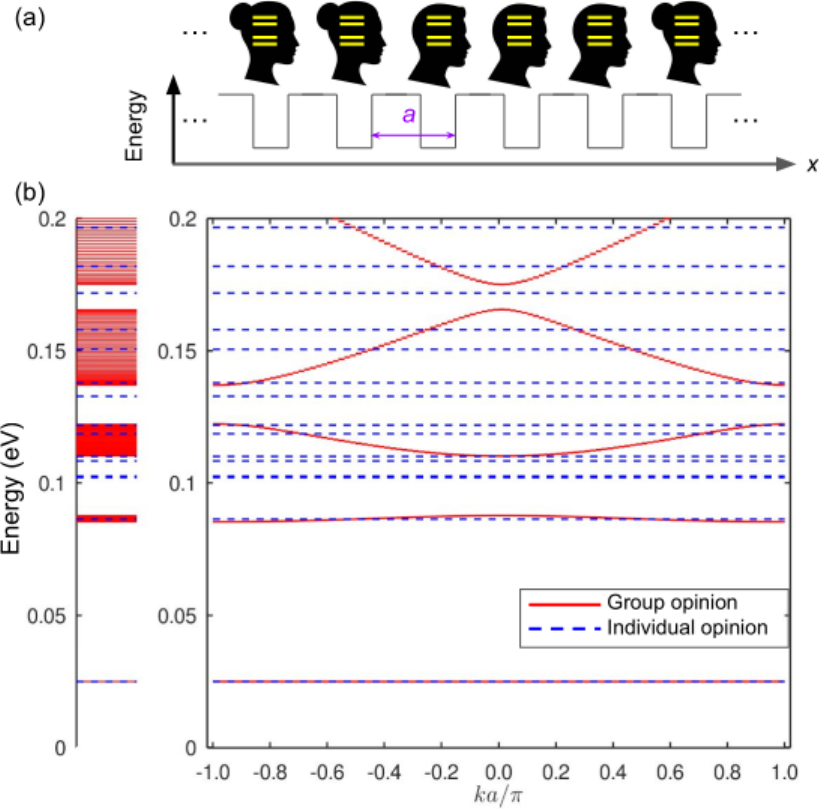}
\caption{Quantised energy level model of opinion polarisation. {\bf(a)}~Social network of like-minded individuals represented as a one-dimensional lattice made of rectangular potential wells (parameter $a$ denotes the period of repetition of the wells). {\bf(b)}~Energy dispersion diagram of the lattice of potential wells (the solid curves in the main panel) and the corresponding energy level structure (the solid lines in the inset). The dashed lines denote the energy levels of a standalone potential well corresponding to an isolated individual. Note that in the group the energy levels split and aggregate, forming bands of continuous energy states, compared with the purely discrete energy levels of the standalone individual. While transitions inside an energy band are readily possible, high energy is required to transition between the bands. This physical property is interpreted as opinion polarisation.}\label{fig:opinion_polarization}
\end{figure}

To establish a physico-mathematical connection between the energy levels and a system of human beliefs, we employ a harmonic quantum oscillator model represented by an electron trapped in a one-dimensional rectangular well, where the energy states are quantised as $E_n \propto (n/L)^2$, being $n=1,2,\dots$ the number of the energy state and $L$ the width of the well \cite{Mes62}. Noting that potential wells of different width $L$ have different allowable energy levels $E_n$, we arrange two or more potential wells into a chain to represent interactions of an individual with social neighbours (Fig.~\ref{fig:opinion_polarization}). We also use the fact that electrons can change energy levels by emitting or absorbing a photon and note that crystalline solids can have energy bands that consist of many closely located discrete energy levels \cite{Kittel}. In the latter case, although electrons are restricted to the band energies, they can assume a continuum of energy values inside a band.

In exact sciences, theoretical models are often based on postulates, as famously exemplified by Einstein's postulates of special relativity \cite{Ein12} and the postulates that underpin Bohr's model of the atom \cite{Kra79}. In general, postulates are accepted to be true without proof since they make a statement that is seen as truth in the framework of the model \cite{Kui16}.

Using this approach, we postulate that in our model a rather regular and sparse energy level pattern corresponds to a polarised (biased) set of beliefs, but the beliefs of an idealised unbiased person or a group of individuals are represented by a continuum of energy values. Other scenarios are also possible, including the case of several continuum state bands separated by large energy gaps that would correspond to division of opinions into sub-groups \cite{Gue21}. An additional relevant discussion can be found in Appendix~A.

We also choose a set of model parameters that has a strict physical meaning. If a person holds deeply polarised beliefs, a significant amount of information would be needed to ameliorate the situation \cite{Rol20}, for example, by means of metacognitive training \cite{Mor19}. Since information has been associated with energy \cite{Adr08, Toy10, Dit14, Vop19, Gos22}, the quantum-mechanical model captures this scenario as a large energy needed for an electron to transition from one energy level to another. However, a set of densely packed energy levels would represent `open-mindedness' because a continuous distribution of energy levels is characterised by a low transition energy, which means that the individual is receptive to new information, ideas and opinions.

To implement the model as a computational code, we numerically solve the Schr{\"o}dinger equation that defines eigenfunctions corresponding to eigenvalues $E$ of the Hamiltonian operator $\hat{H}$ as \cite{Mes62}
\begin{equation}
  \label{eq:SE}
  \hat{H}\psi({\bf r}) \equiv \left[-\frac{\hbar^2}{2m}\Delta + V({\bf r})\right]\psi({\bf r}) = E\psi({\bf r})\,, 
\end{equation}
where $\hbar$ is Plank’s constant, $\Delta$ is the Laplacian operator, $m$ is the mass of the electron and $V({\bf r})$ is the scalar potential. We employ a one-dimensional finite-difference method where Eq.~(\ref{eq:SE}) is discretised along the coordinate $x$ so that $V(x)$ is represented as a vector of $N_x$ equally spaced points with a step $h_x$ \cite{Hal22}. Using a second-order central finite-difference scheme, we obtain
\begin{equation}
  \label{eq:deriv2}
  \psi''(x_i) \approx \frac{1}{h_x^2}\left[\psi(x_{i-1})-2\psi(x_{i})+\psi(x_{i+1})\right]\,. 
\end{equation}
The substitution of Eq.~(\ref{eq:deriv2}) to Eq.~(\ref{eq:SE}) gives rise to a finite-difference version of the Schr{\"o}dinger equation written in a matrix form. We employ Floquet periodic boundary conditions $\psi(x_{0})=\psi(x_{Nx})\exp(-jka)$ and $\psi(x_{Nx+1})=\psi(x_{1})\exp(jka)$, where $j$ is the unit imaginary number, $k$ denotes the wavevector and $a$ is the period of repetition of the profile of $V(x)$ (for an isolated potential well $\psi(x_0)=\psi(x_{Nx+1})=0$). 

The resulting matrix equation is solved using a standard procedure {\tt eigs} of MATLAB/Octave software to find the eigenvalues $E$ for each discrete value of the wavevector $k$ corresponding to the first Brillouin zone. The energy level diagrams plotted alongside the dispersion diagrams are obtained integrating the values of $E$ for all values of $k$ taken into account in the calculation.

\subsection{Opinion Polarisation in Social Networks\label{OpinPol}}
We first use our model to capture the phenomenon of opinion polarisation in a social network. Then, we explain the technical aspects of the model and demonstrate a relationship between the physics and dynamics of social networks, relaxing the need for the readers to understand physics-specific terms.  

When two individuals do not interact and are not influenced by the news media, the systems of their beliefs, represented by the energy levels, remain unchanged because the corresponding wave functions---probability waves that govern the motion of the electron and are described by the Schr{\"o}dinger equation \cite{Mes62}---do not overlap. However, as a result of information exchange between these individuals, the potential wells move closer one to another and the wave functions overlap, thereby forcing the energy levels to adjust and split such that they remain unique \cite{Kittel}.

Figure~\ref{fig:opinion_polarization} shows the result of a simulation of the opinion polarisation in a social network, where we form a social network of like-minded individuals, i.e.~individuals whose systems of beliefs are represent by the same discrete energy level systems. We take into account a large number of members of the social group and represent them as a periodically arranged sequence (one-dimensional lattice) of identical potential wells (Fig.~\ref{fig:opinion_polarization}a). 

The main panel of Fig.~\ref{fig:opinion_polarization}b shows the calculated energy dispersion diagram (the solid curves) alongside the energy levels of a standalone single potential well (the dashed lines). The former represent the group opinion formed in a social network of like-minded individuals but the latter corresponds to the opinion of one individual considered separately from social neighbours. The right inset to Fig.~\ref{fig:opinion_polarization}b shows the discrete energy levels corresponding to the group (the solid lines) and the isolated individual (the dashed lines). In this calculation, the depth and width of the wells are 0.1\,eV and 10\,nm, respectively (the meaning of these parameters will be revealed below).

We can see that, by virtue of the laws of quantum mechanics \cite{Kittel}, in our model of the social network the original discrete energy levels corresponding to the individuals split and self-arrange, forming several allowed energy bands separated by band gaps (regions of forbidden energies \cite{Kittel}). According to the postulates of our model, the formation of energy bands corresponds to the polarisation of opinions: the opinions of the group participants is more likely to remain withing the energy bands but changes of opinion are less likely because a high energy is required for an interband transition. 

The presence of the fundamental effect of energy splitting differentiates our model from any previous physics-inspired models of social interaction (for a more detailed discussion see Appendix~A). Indeed, in the Sznajd model \cite{Szn00} and its modifications \cite{Red19}, the interaction between two neighbours changes the opinions of their respective neighbours. However, the opinions of these two particular neighbours are assumed to be unchanged. A similar assumption is also made in the other models \cite{Gal05, Ort13, Pak13}.

However, generally speaking, making this assumption contradicts the accepted philosophical vision of the dialogue as a means of change \cite{Ros18, Put22}. In fact, dialogue entails such quality relationships between individuals as mutuality, responsibility, engagement and acceptance, inevitably resulting in the evolution of personal beliefs and views \cite{Kov11}. While such a change may materialise in a long-term perspective, thus making it possible to neglect it in some social model situation such as the prediction of political election exit polls \cite{Lar13}, it should be taken into account in a model of opinion polarisation in social media where dialogue plays an important role \cite{Jov18}.  

In the calculation used to produce the data plotted in Fig.~\ref{fig:opinion_polarization}b, we denoted the period of repetition of the potential wells as $a$ and applied Floquet periodic boundary conditions that are often employed in solid state physics to approximate a large system using a small part called the unit cell \cite{Kittel}. We also used the concept of the reciprocal lattice that represents the Fourier transform of the periodic lattice of potential wells in the physical space (the $x$-coordinate in Fig.~\ref{fig:opinion_polarization}a) and that exists in the reciprocal space of wavevector $k$. According to the convention adopted in the field of solid state physics, we plotted the energy dispersion diagram as a function of the normalised wavevector $ka/\pi$ in the range of values corresponding to the first Brillouin zone that defines the unit cell in the reciprocal space \cite{Kittel}.
\begin{figure}[t]
\centering
\includegraphics[width=.85\linewidth]{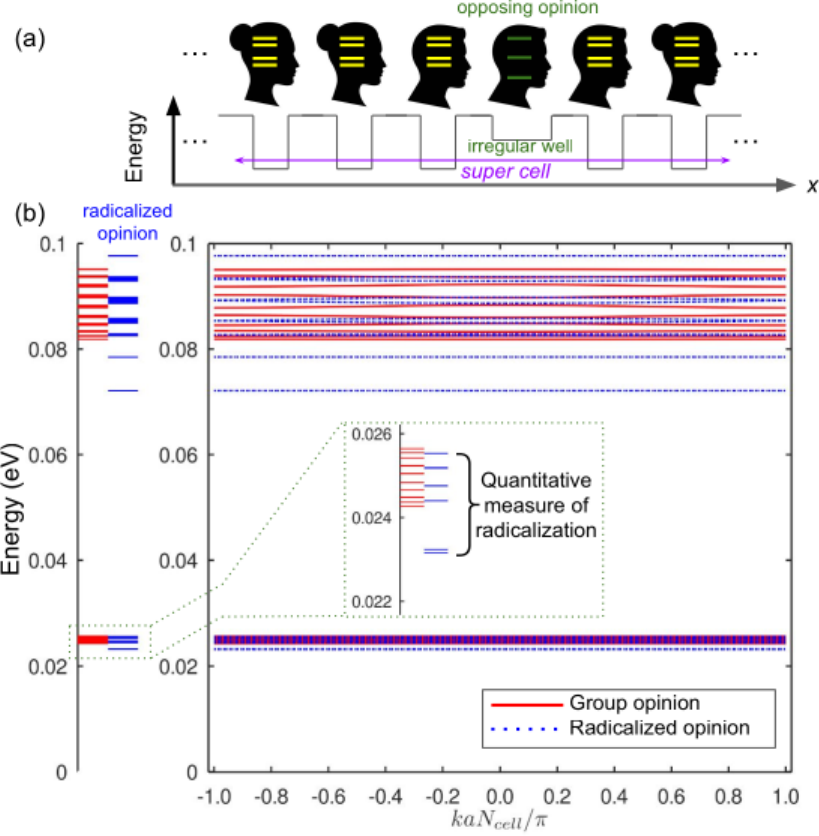}
\caption{Example of opinion polarisation. {\bf(a)}~Like-minded individuals are exposed to an opposing opinion, represented by a one-dimensional lattice of identical potential wells with a `defect' lattice node given by an irregular well. A super cell approach described in the main text is employed in the calculation. {\bf(b)}~Main panel:~Energy dispersion diagram corresponding to the group opinion before the exposure to an opposing opinion (the solid curves) and after it (the dotted curves). The respective energy level structures are shown in the left inset. Note that the exposure to the opposing opinion results in addition of new discrete energy levels and splitting of the existing ones, which can be seen both in the right and central insets and which is interpreted as opinion polarisation. The degree of opinion polarisation can be estimated computing the energy difference between the lowest and highest energy levels, as shown in the central inset.}\label{fig:opinion_radicalization}
\end{figure}

Whereas the discussion above employs the terminology adopted in physics, we can establish a link between the physical concepts and the concepts used in data science and studies of social networks. Indeed, studies of social networks often exploit the notion of periodicity and rely on the Fourier transformation of data \cite{Ric19, Bas21}, highlighting an important role of the reciprocal space in analysis of big datasets \cite{Mcc12, Vas15}. Hence, the readers who do not wish to use the physical language can also comprehend our model using the terminology adopted in the fields of graph signal processing \cite{Bas21}, data science \cite{Ric19} or mathematical sociology \cite{Mcc12}. Moreover, we presented energy level plots alongside the dispersion diagrams, helping the readers understand the mainstream discussion without the need to use the concepts of wavevector and Brillouin zone. We also note that the depth and width of the wells used in our model can be made nondimensional, enabling the users of the model to employ a system of measure that is typical in their field of research. For example, the spacing between the discrete energy states can be related to the social distance \cite{Par19} or opinion distance in social networks \cite{Ant18, Kui23, Nug23}.
\begin{figure}[t]
\centering
\includegraphics[width=.85\linewidth]{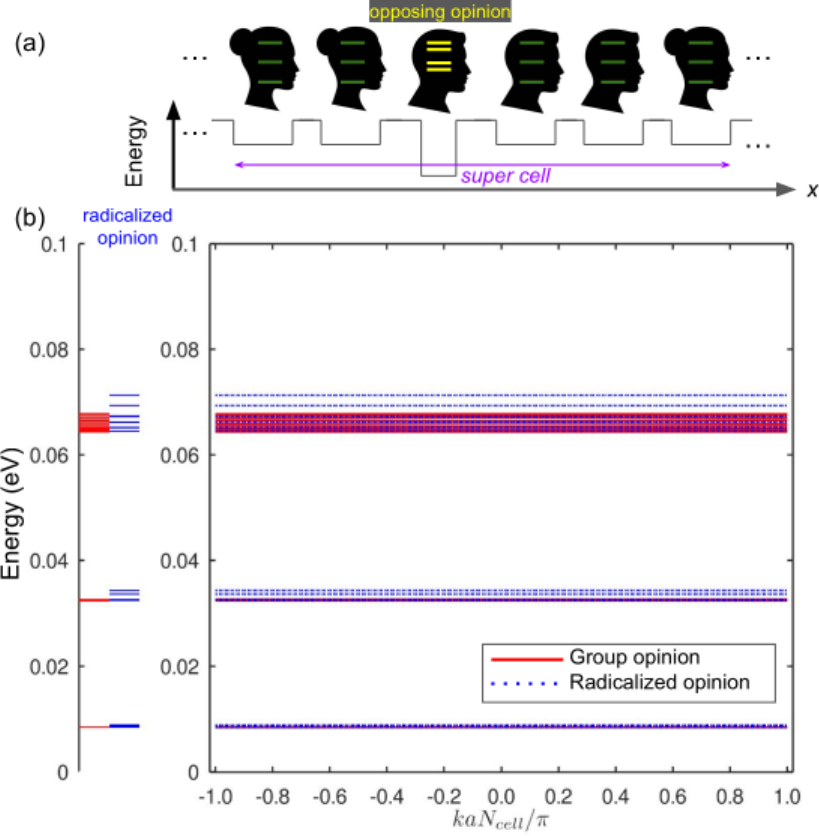}
\caption{Example of opinion polarisation in a reverse scenario with respect to Fig.~\ref{fig:opinion_radicalization}a. {\bf(a)}~Like-minded individuals are exposed to an opposing opinion, represented by a one-dimensional lattice of identical potential wells with a `defect' lattice node. Unlike in Fig.~\ref{fig:opinion_radicalization}a, the irregular well is deep and narrow but the majority opinion wells are shallow and wide. {\bf(b)}~Main panel:~Energy dispersion diagram corresponding to the group opinion before the exposure to an opposing opinion (the solid curves) and after it (the dotted curves). The respective energy level structures are shown in the left inset. Even though the exposure to the opposing opinion results in addition of new discrete energy levels and splitting of the existing ones as in Fig.~\ref{fig:opinion_radicalization}b, these processes are less pronounced thereby indicating a lower degree of opinion polarisation with respect to the scenario modelled in Fig.~\ref{fig:opinion_radicalization}.}\label{fig:opinion_radicalization_reverse}
\end{figure}

\subsection{Opinion Radicalisation}
Now we showcase the ability of the model to capture the backfire effect and opinion radicalisation in social networks. As an example, we refer to a study demonstrating that exposure to opposing views on social media can radicalise political views \cite{Bai18}. In the cited paper, members of the Republican and Democratic parties received financial compensation to follow officials and opinion leaders with opposing political views. As a result of the experiment, compared with the control group, both Republican and Democrat participants developed more radicalised views with respect to their respective traditional party positions \cite{Bai18}. Interestingly, the degree of opinion radicalisation was different among the two groups: the Republican participants showed substantially more conservative views but the Democrat participants exhibited just a slight but statistically significant increase in liberal attitudes (Fig.~\ref{fig:litexamples}, top panel). 

The corresponding model situation is presented in Fig.~\ref{fig:opinion_radicalization}a, where persons with the same systems of beliefs, represented by regular potential wells, are exposed to an opposing opinion expressed by a person represented by an irregular potential well. Unlike the regular potential wells that are 0.1\,eV deep and 10\,nm wide, the irregular well is 0.07\,eV deep and 20\,nm wide (we remind that the parameters of the wells are chosen empirically and they are meaningful only in the framework of the postulates that underpin our model). Subsequently, the allowed energy levels of an individual irregular well taken alone are significantly shifted with respect to those of an individual regular well. To account for the presence of the irregular well in an otherwise periodic sequence of identical regular wells, we employ a super cell computational approach that is often used to model a crystallographic defect, an interruption of the regular patterns of arrangement of atoms in crystalline solids \cite{Kittel}.

While the super cell approach appears to be unique to the domains of physics and mathematics, we show that it can easily be adopted in models of social networks. An infinitely large, idealised network consisting of identical individuals can be mathematically represented as one individual (the unit cell) that self-repeats an infinite number of times. A more realistic network will consists of different individuals. However, a careful analysis of such a network is likely to reveal the existence of sub-groups of similar individuals \cite{Van19}. Considering such sub-groups as super cells, we can create a model of the network where the super cells repeat as many times as needed to reproduce the structure of the network. We can further extend this idea by manipulating the opinions of individuals within a sub-group, i.e.~within a super cell. For instance, we can require one individual within a sub-group to suddenly change their opinion for an opposite one. This action is equivalent to the introduction of a crystallographic defect, which is the reason why we adopt the physical term `defect' in the discussion of our model.         

In the main panel of Fig.~\ref{fig:opinion_radicalization}b, the solid lines denote the energy dispersion diagram corresponding to a polarised opinion in a group of like-minded individuals, which corresponds to the control group where the irregular well is replaced by a regular one. The dotted lines plot the energy dispersion in the case of exposure to an opposing opinion. The corresponding energy bands are shown in the inset. For the sake of presentation, we limit the energy range of interest to $0\dots0.1$\,eV. For the readers interested in the physical and mathematical aspects of the model, we also note that the introduction of a super cell consisting of $N_{cell}$ potential wells effectively folds the Brillouin zone, modifying the energy dispersion curves and requiring an adjustment of the wavevector normalisation as $kaN_{cell}/\pi$ compared with Fig.~\ref{fig:opinion_polarization}b.

We can see that the exposure to an opposing opinion changes the energy level structure because a defect in an otherwise perfectly periodic lattice of potential wells produces a series of sparse energy levels. Since we stipulated that a sparse energy level pattern corresponds to a polarised set of beliefs, we conclude that our model reproduces the backfire effect. 

According to the data presented in Fig.~\ref{fig:litexamples}, the radicalisation should also occur in a scenario that is reverse to Fig.~\ref{fig:opinion_radicalization}a: the majority opinion becomes represented by the potential well considered in Fig.~\ref{fig:opinion_radicalization}a as irregular but the formerly regular well serves as the `defect'. We simulate this scenario using a super cell consisting of 0.07\,eV deep and 20\,nm wide potential wells containing a 0.1\,eV deep and 10\,nm wide irregular well (Fig.~\ref{fig:opinion_radicalization_reverse}a), which is the structure that is topologically opposite to that shown in Fig.~\ref{fig:opinion_radicalization}a.

The result of the corresponding calculation is presented in Fig.~\ref{fig:opinion_radicalization_reverse}b, where we observe that the exposure to an opposite view leads to the formation of sparse discrete energy levels, i.e.~to a radicalisation of the group opinion. We can see that the sparse energy levels in Fig.~\ref{fig:opinion_radicalization_reverse}b are quantitatively different from those in Fig.~\ref{fig:opinion_radicalization}b, which indicates that the levels of opinion radicalisation corresponding to the results presented in these figures are also different. Below we will show that the difference between the results in Fig.~\ref{fig:opinion_radicalization_reverse}b and Fig.~\ref{fig:opinion_radicalization}b correlates with the data presented in Fig.~\ref{fig:litexamples}.

In the work \cite{Bai18} that presents the asymmetric radicalisation data used in the top panel of Fig.~\ref{fig:litexamples}, the degree of radicalisation was studied using the Rubin casual model of potential outcomes \cite{Imb15}, resulting in a quantitative liberal/conservative scale based on experimental data obtained from a mainstream social network. The energy states plotted in the insets to Fig.~\ref{fig:opinion_radicalization}b and Fig.~\ref{fig:opinion_radicalization_reverse}b are analogous to the scale used in Ref.~\cite{Bai18}. To provide a measure of radicalisation, in our model we first calculate the difference between the highest and lowest `Group opinion' energy states and then compare the resulting values with the difference between the respective `Radicalised opinion' energy states. Applying this procedure to the energy bands at approximately 0.022\,eV in Fig.~\ref{fig:opinion_radicalization}b and 0.008\,eV in Fig.~\ref{fig:opinion_radicalization_reverse}b, we establish that the opinion radicalisation in Fig.~\ref{fig:opinion_radicalization} is a factor of 1.5 stronger than in Fig.~\ref{fig:opinion_radicalization_reverse}.

This result is close to the ratio of the experimental values for Non-Religious/Religious (Fig.~\ref{fig:litexamples}, bottom panel) and consistent with the radicalisation ratio for Democrats/Republicans (Fig.~\ref{fig:litexamples}, top panel). Thus, although further model adjustments are required to obtain more accurate results, it is plausible that the model may predict at least approximate radicalisation data for pro-vaccine and pro-climate change groups (those data are not available as explained in the caption to Fig.~\ref{fig:litexamples}) using the data available for anti-vaccine groups and climate change sceptics. In particular, the model confirms an intuitive conclusion that can be made observing Fig.~\ref{fig:litexamples}: social groups that share more liberal views experience a mild radicalisation due to the exposure to conservative views compared with a considerable radicalisation of conservative social groups exposed to liberal views.      

\section{Discussion}
The model developed in this paper adopts a new approach that has not been used in the previous quantum models of decision-making and opinion formation. Since this work is a joint effort of a physicist working on the social aspects of AI and an expert in psychology, behavioural science, economics and decision-making, it presents a integrative view on the quantum-mechanical models of opinion formation and polarisation compared with the previous works written by experts in the traditional fields of research. In particular, this paper combines the physical concept of the electronic band structure that represents the energy levels with intricate psychological effects, elaborating on the previous suggestions to describe the mental states as discrete energy levels \cite{Aer22, Aer22_1} and also bridging a gap between psychological studies and interdisciplinary academic works that broadly contribute to the emergent research topics of quantum mind and socio-physics. Since this paper targets the readers specialised in different research areas ranging from psychology and decision-making to data science, mathematics and physics, in this section we use a field-of-research neutral language to summarise the main features of the proposed model, also demonstrating the place of the model in the taxonomy of socio-physical approaches to opinion formation in social networks.      

\subsection{The Origin of Quantum Advantage in Models of Social Media}
Socio-physical models based on classical physical effects \cite{DeG74, Szn00, Gal05, Cas09, Red19, Jus22, Gal22_1} have become an important topic of mainstream scientific research and found practical applications outside academia \cite{Gal17, Gal18}. Thus, one of the main questions of this paper is what in general makes a quantum-mechanical model attractive for the analysis of opinion polarisation in social networks.

To address this question, let us compare a classical digital computer with a quantum computer \cite{Nie02}. Similarly to an on-off light switch, a bit of a digital computer is always in one of two physical states corresponding to the binary values `0' and `1'. However, a quantum computer uses a quantum bit (qubit) that can be in the states $|0 \rangle = \icol{1\\0}$ and $|1 \rangle = \icol{0\\1}$. While these states are analogous to the `0' and `1' binary states of the digital computer, a qubit also exists in a superposition of the states $|0 \rangle$ and $|1 \rangle$ expressed as $|\psi\rangle = \alpha |0 \rangle + \beta |1 \rangle$ with $|\alpha|^2 + |\beta|^2 = 1$.

\begin{figure}
 \includegraphics[width=0.5\textwidth]{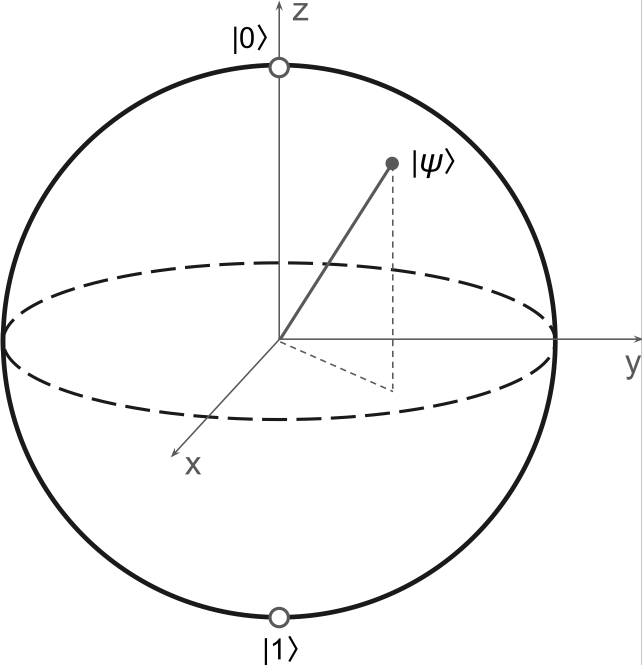}
 \caption{Illustration of a projective measurement of a qubit $|\psi\rangle$ using the Bloch sphere.\label{Fig_qubit}}
\end{figure}

Computational algorithms based on measurements of a qubit are exponentially faster than any possible deterministic classical algorithm \cite{Nie02}. This advantage can be illustrated using the concept of the Bloch sphere (Fig.~\ref{Fig_qubit}) that depicts to the multitude of the qubit states that can be used to conduct a calculation. When a quantum measurement is done \cite{Mes62, Nie02}, a closed qubit system interacts in a controlled way with an external system, revealing the state of the qubit under measurement. Using the projective measurement operators $M_0 = |0\rangle\langle0|$ and $M_1 = |1\rangle\langle1|$ \cite{Nie02}, the measurement probabilities for $|\psi\rangle = \alpha |0 \rangle + \beta |1 \rangle$ are $P_{|0\rangle} = |\alpha|^2$ and $P_{|1\rangle} = |\beta|^2$, which means that the qubit will be in one of its basis states. Visually, the measurement procedure means that the qubit may be projected on one of the coordinate axes of Fig.~\ref{Fig_qubit}.

Thus, it has been demonstrated that quantum mechanics can model the human mental states better than any existing classical model \cite{Pot09, Bus12, Ben18, Pot22}. Indeed, referring to the classical models where the opinion states can be either `0' and `1' \cite{Szn00, Red19}, in Fig.~\ref{Fig_qubit} we can see that the quantum mechanics enables us to describe human opinions using the multitude of the possible qubit states. In particular, while in the classical models the two allowed states can be limited to `Agree' and `Disagree' \cite{Szn00, Red19}, in a quantum-mechanical model we can have a spectrum of opinions that gradually vary from `Agree' and `Disagree'. More intriguingly, a quantum-mechanical model of cognition is not constricted by any technical limitations of hardware physical systems that realise qubits. This means that in a model we can unleash the full potential of the qubit states (as well as of more complex states as discussed in Appendix~A) to describe complex human mental states \cite{Khr06, Atm10, Bus12}.

Further evidence of the superiority of quantum-mechanical models over the classical physical ones can be provided considering the recent advances in the field of quantitative finance where the methods of quantum physics play an increasingly important role \cite{Baa20, Her23}. Historically, the financial markets have been strongly influenced by the emotions of traders \cite{Nor09}. Therefore, it has been established that the processes that underpin the formation of opinions of traders and the formation of opinion in social networks are essentially similar and intertwined \cite{Ped22, Hir23} and that their analysis benefits from the richness of the quantum states used in respective quantum quantitative finance models \cite{Baa20, Her23}.

In fact, both financial markets \cite{Fab22} and human brain \cite{Mck94, Kor03} are dynamical systems. A dynamical system is a mathematical or physical system whose evolution can be described by differential equations that involve time derivatives. Subsequently, financial markets can be described by an appropriate differential equation that, for example, may correspond to a classical harmonic oscillator \cite{Gar20}. Yet, it was demonstrated that a quantum harmonic oscillator governed by the Schr{\"o}dinger equation producing quantised eigenfunction solutions---essentially it is Eq.~(\ref{eq:SE}) of our model---serves as a better model of financial markets \cite{Ahn18}. Importantly, as with the model presented in this paper, the quantum oscillator financial model outperforms the classical oscillator financial models due to the quantum effects resulting in the formation of discrete energy levels. The same conclusion has been drawn in the relevant quantum mind studies where quantum oscillator models demonstrated significant advantages over the classical models \cite{Bus12, Mak24_illusions}.

Finally, the fundamental principles of quantum mechanics used in this paper have also been employed in the recent work \cite{Shi23} that demonstrated that quantum processes can help find the conditions suitable for making conflict-free joint decisions between two individuals. Although the authors of the cited paper considered the quantum processes of photon entanglement and interference of orbital angular momentum, at the fundamental level they reached similar conclusions. In particular, they revealed the importance of the symmetry of quantum system for decision-making modelling. The relevance of symmetry to the model presented in this paper is discussed in Appendix~A.  

\subsection{Quantum-Mechanical Model vs. Statistics-Based and Data Science-Driven Approaches}
Now we compare the quantum-mechanical model with the traditional statistics-based and data science-driven approaches. As with the established classical physics-based models of opinion formation in social media \cite{Szn00, Gal05, Red19, Gal22}, the present quantum-mechanical model conceptually differs from the statistics-based and data science-driven methodologies adopted in the fields of psychology, econometrics and decision-making \cite{Imb15}. Specifically, it provides a robust insight into the origin of the radicalisation effect without relying on large data sets obtained in complex and time-consuming experiments involving large groups of population.

Indeed, a typical single run of a computer program that implements the model presented in this paper is about 10\,s, which enables the user to gain an understanding of the processes underpinning the radicalisation in a very short period of time compared with the time needed to collect, process and analyse experimental data (see the works cited in the caption to Fig.~\ref{fig:litexamples}). While the information about the time required to collect and process data is not readily available in the literature, in our own relevant work \cite{Mak24_preference} we demonstrated that it takes approximately one hour of CPU time of a workstation computer for a quantum model to reproduce empirical psychological datasets obtained as a result of experiments conducted over several months.     

As a next development stage, the agreement between the outcomes produced by our model and the experimental results can be increased using the fact that the potential well profiles capturing an intricate system of human beliefs can be arbitrary complex. That is, the potential wells may not necessarily be rectangular but can assume a parabolic or a triangular shape or their combinations. To automatise the search for a suitable profile, one can use machine learning techniques that have been employed to optimise the structure of photonic crystals, periodic dielectric structures that share some physical properties with a periodic array of potential wells \cite{Chr20}.

\subsection{Integration with Bot Detection Systems and Decision-Making Software}
Bot detection is the process of identifying and distinguishing between automated bots and human users \cite{Ora20, Hay23}. Bots can be detected using algorithms based on behavioural analysis, challenge-response authentication systems and machine learning techniques \cite{Ora20}. Nevertheless, despite a constant progress in the domain of cyber security, bots remain a serious threat since they can shape the public opinion by spreading fake news and serving as an instrument for cyberbullying and harassment \cite{Ora20, Hay23}.

At present, many bot detection systems analyse user behaviour patterns, such as page navigation habits, to detect anomalies that may indicate bot activity. The model proposed in this paper offers the opportunity to extend the capability of bot detection systems by modelling the way humans think while they use social networks. Comparing such models with the decision-making patterns of users of social networks, a bot detection system may differentiate a real user from a bot. This functionality should be especially useful since the developers of bots employ artificial intelligence systems that learn to accurately reproduce certain behaviour patterns of social network users \cite{Hay23}. However, even the most sophisticated bot system cannot reproduce human thoughts yet, thereby revealing the synthetic nature of the bot when its behaviour is benchmarked against a quantum digital twin of a human.

A quantum digital twin can also be used in decision-making software that helps business leaders make their decisions based on a large number of factors, including the presence in social media and feedback received from them. According to a recent report \cite{KPMG}, 41\% of senior leaders in Australia make their decisions on instinct alone and 81\% report their business suffered the consequences of such decisions. Similar decision-making distress also affects leaders in politics, police force, secret intelligence and army, causing them to question or regret decisions they have made. The development of quantum models of human behaviour aligns with the vision expressed in the report \cite{KPMG}: a judicious combination of artificial intelligence systems with advanced mathematical models should help extend the ability of a human mind to predict decision outcomes and de-risk operational transformation.    

\section{Conclusion}
This paper demonstrates the potential of a quantum-mechanical model to illustrate, interpret and explain the dynamics of opinion polarisation and radicalisation in social networks. We have shown that the structure of energy levels of a quantum system, specifically of a chain of potential wells, can effectively model social dynamics such as interactions between individuals and groups.

Our results indicate that the phenomenon of opinion polarisation can be represented by the formation of continuous energy bands and interband gaps when like-minded individuals interact within a social network. These energy bands represent the polarised groups, with transitions between them requiring significant energy. The ability of this model to capture the experimental results obtained using the established social psychology methods is a promising indication of its predictive capacity.

Moreover, we extended this model to capture the effects of exposing like-minded groups to opposing views, demonstrating that this situation can lead to the creation of new, sparse energy levels, symbolising the radicalisation of the group's opinion. This is interpreted as the backfire effect, a behaviour observed in real-world scenarios and supported by empirical studies. We also showcased how variations in potential well profiles can influence the degree of opinion radicalisation, corroborating with the observation that different groups might react differently to the exposure to opposing views.

Utilising a novel approach of conceptualising social interactions, this model provides a theoretical basis for understanding and predicting social dynamics in intricate scenarios. Although this work primarily employs a one-dimensional model, it opens the door to future research extending the model to higher spatial dimensions and considering more complex potential well profiles. In conjunction with the ability of machine learning techniques to synthesise profiles that capture intricate systems of human beliefs, emotions and brain activity, this property enables us to further refine the model and better represent and anticipate the multifaceted reality of social network interactions.

\vspace{6pt}

\authorcontributions{I.S.M.: conceptualisation, data curation, formal analysis, investigation, methodology, writing—original draft; writing—review and editing; G.P.: conceptualisation, data curation, formal analysis, investigation, methodology, writing—original draft; writing—review and editing.}

\funding{The authors did not receive any funding to support this project.}

\institutionalreview{Not applicable.}

\informedconsent{Not applicable.}

\dataavailability{This article has no additional data. The computational code used in this work can be found at \url{https://github.com/IvanMaksymov/OpinionPolarisation}}


\conflictsofinterest{The authors declare no conflict of interest.}

%
%



\appendixtitles{no} 
\appendixstart
\appendix
\section[\appendixname~\thesection]{Additional Discussion and Future Research Directions}
The electron spin is a purely quantum-mechanical concept \cite{Mes62, Kittel}. However, as discussed in more detail in \cite{Mak24_gender, Mak24_preference}, certain properties of the spin can be described using a classical magnetic dipole, making it possible to use the spin states $\uparrow$ and $\downarrow$ in the classical socio-physical models \cite{Szn00, Red19}. At the conceptual level, these states may be compared with the basis $|0\rangle$ and $|1\rangle$ states shown in Fig.~\ref{Fig_qubit}. 

In an idealised scenario, a classical physical magnetic dipole model of opinion formation may involve two interacting particles---individuals $A$ and $B$---that can assume one of the two allowable states $\xi^{\prime}$ and $\xi^{\prime\prime}$ defined at any time by the position and momentum $\xi_{A}\equiv({\bf r}_{A}, {\bf p}_{A})$ and $\xi_{B}\equiv({\bf r}_{B}, {\bf p}_{B})$. The corresponding Hamiltonian function is $\Hamiltonian(\xi_{A}, \xi_{B}) = {\bf p}_{A}^2/(2m) + {\bf p}_{B}^2/(2m) + V(|{\bf r}_{A}-{\bf r}_{B}|)$, where $m$ is the mass of the particles and $V({\bf r})$ is the profile of potential \cite{Mes62}. Since $\Hamiltonian(\xi^{\prime}, \xi^{\prime\prime}) = \Hamiltonian(\xi^{\prime\prime}, \xi^{\prime})$, we can conclude that in the classical model the common opinion state of a group consisting of two individuals is known to within the permutation of the individuals. This ambiguity should not represent any difficulty in modelling the opinion dynamics. However, it does not enable one to differentiate the impact of the views of $A$ on the views of $B$ and vice versa, which is disadvantageous for modelling the asymmetry of opinion radicalisation.

The treatment of the two-individuals system using methods of quantum mechanics produces a more complex physical picture. Assuming that the states of $A$ and $B$ are $\psi^{\prime}({\bf r})$ and $\psi^{\prime\prime}({\bf r})$, we can show that the observation of the interaction between $A$ and $B$ does not permit to decide whether the group opinion is in the state $\psi({\bf r}_{A}, {\bf r}_{B}) \equiv \psi^{\prime}({\bf r}_{A})\psi^{\prime\prime}({\bf r}_{B})$ or $\hat{\psi}({\bf r}_{A}, {\bf r}_{B}) \equiv \psi^{\prime\prime}({\bf r}_{A})\psi^{\prime}({\bf r}_{B})$. This situation is called exchange degeneracy and it originates from the fact that the functions $\psi$ and $\hat{\psi}$ are both eigenfunctions corresponding to a set values produced by the measurement \cite{Mes62}. Subsequently, any linear combination $\alpha\psi+\beta\hat{\psi}$ with $|\alpha|^2+|\beta|^2=1$ can represent the state of the system.

Quantum mechanics removes the exchange degeneracy introducing a symmetrisation postulate that fixes the coefficients $\alpha$ and $\beta$ of the linear combination of the states $\psi$ and $\hat{\psi}$ \cite{Mes62}. Therefore, the states of the system consisting of individuals $A$ and $B$ are necessarily either all symmetrical ($\alpha=0$, $\beta=1$) or all antisymmetrical  ($\alpha=1$, $\beta=0$) in the permutation of $A$ and $B$. The symmetrisation applies to a system of $N$ particles/individuals \cite{Mes62} and it can be used to capture the effect of asymmetric opinion polarisation.

In the proposed model, the corresponding states are antisymmetrical and they are governed by the Pauli exclusion principle, according to which no two electrons can occupy the same quantum state \cite{Mes62}. Thus, in light of the illustration in Fig.~\ref{fig:social_energy_levels}, by analogy with solid state physical systems \cite{Kittel} each orbital labelled by the quantum number $n$ can accommodate two individuals, one with with the state $\uparrow$ and another one with the state $\downarrow$. Moreover, when the orbitals overlap, the Pauli principle prevents multiple state occupancy, resulting in a change in the total energy of the system \cite{Kittel}. While this physical picture does not necessarily reflect the actual human mental states (in fact, models that emulate the actual mental states may suffer from fundamental drawbacks \cite{Tad16}), its integration into the model of social opinion dynamics adds novel degrees of freedom that enable a deeper understanding of complex human decision-making processes \cite{Bus12, Aer22, Aer22_1}.
\begin{figure}
 \includegraphics[width=0.99\textwidth]{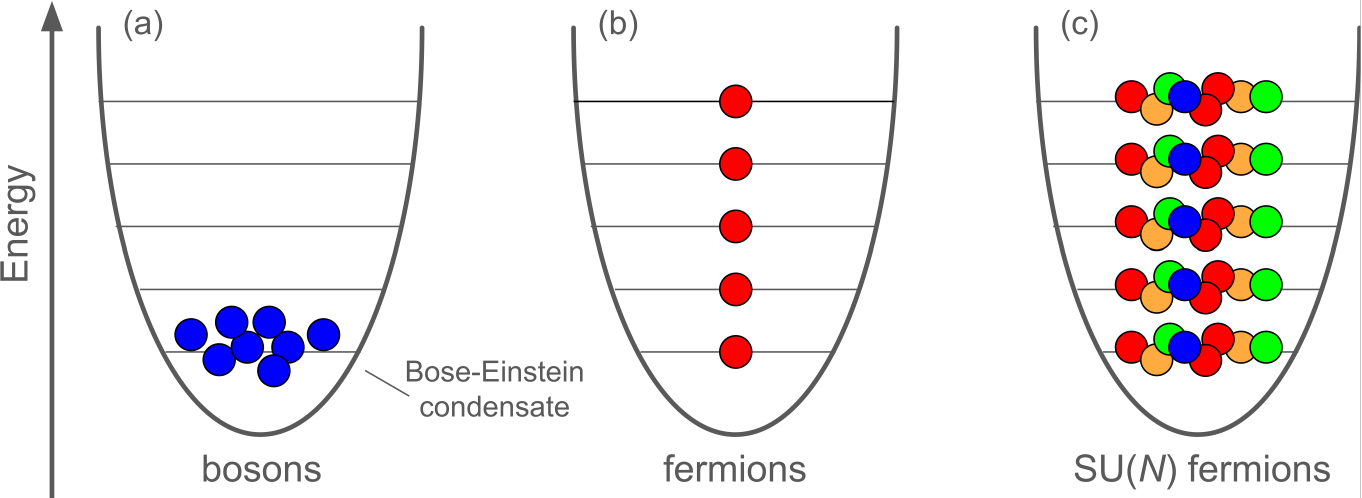}
 \caption{{\bf(a)}~Degenerate bosons, {\bf(b)}~fermions and {\bf(c)}~SU($N$) fermions. Unlike bosons that can occupy the same energy level at low temperatures, fermions separate into different energy levels. However, SU($N$) fermions can have $N$ particles per energy level. In each level, each particle has $(N-1)$ distinct neighbours that strongly interact one with another.\label{Fig_bosons}}
\end{figure}

The particles with antisymmetrical states are called fermions \cite{Mes62}. Unlike fermions, bosons---particles with symmetrical states---are not subject to the Pauli exclusion principle, i.e.~two or more bosons can occupy the same quantum state (Figure~\ref{Fig_bosons}). If we have two experimental setups, one containing a gas of bosons but another one contains a gas of non-interacting fermions, as the temperature decreases to absolute zero, the gas of bosons collapses, forming a Bose-Einstein Condensate (Fig.~\ref{Fig_bosons}a). However, fermions cannot reach this state since they cannot occupy the same quantum state: they occupy the discrete energy levels as depicted in Fig.~\ref{Fig_bosons}b, where the energy of the highest occupied quantum state is called the Fermi energy. 

Thus, we anticipate that the incorporation of the quantum physical principles that underpin the properties of fermions and bosons into a model of opinion formation in social media will enables us to understand and predict even more complex social interactions. Interestingly, a similar idea was expressed in the field of quantum modelling of financial markets \cite{Ras19}, which additionally speaks in favour of the correctness of our discussion that established a link between the quantum models of social networks and the quantum models used in the finance sector. Yet, intriguingly, similar models have been shown to correctly describe the arrangement of people in a concert and cars in a parking lot \cite{Per20}. Taken together, these applications demonstrate a significant potential of quantum models to capture complex human decision-making patterns.

Finally, utilising the group theory---the mathematical framework for describing symmetry properties of quantum mechanical systems \cite{Tun85}---and drawing on the recent advances in the field of quantum physics \cite{Miy10, Tai22}, we suggest a new research direction for further development of quantum-mechanical models of human cognition and decision-making. A recent work demonstrated that some real-life materials can have $N$ distinct states, each of which have the same properties due to the symmetry \cite{Son20}. Such physical systems are called SU($N$) symmetric \cite{Tun85, Miy10}. In particular, SU($N$) fermions can have $N$ particles per energy level and, in each level, each particle has $(N-1)$ distinct neighbours that strongly interact one with another (Fig.~\ref{Fig_bosons}c). Taking this property into account in a model of social dynamics should dramatically increase the ability of the model presented in this paper, introducing additional degrees of freedom that can be used to represent the human mental states.


\begin{adjustwidth}{-\extralength}{0cm}
\reftitle{References}


\externalbibliography{yes}
\bibliography{biblio}

\begin{thebibliography}{999}

\bibitem[Ma \em{et~al.}(2019)Ma, Dixon, and Hmielowski]{ma2019psychological}
Ma, Y.; Dixon, G.; Hmielowski, J.D.
\newblock Psychological reactance from reading basic facts on climate change:
  The role of prior views and political identification.
\newblock {\em Environ.~Commun.} {\bf 2019}, {\em 13},~71--86.

\bibitem[Dixon \em{et~al.}(2019)Dixon, Bullock, and Adams]{dixon2019unintended}
Dixon, G.; Bullock, O.; Adams, D.
\newblock Unintended effects of emphasizing the role of climate change in
  recent natural disasters.
\newblock {\em Environ.~Commun.} {\bf 2019}, {\em 13},~135--143.

\bibitem[Druckman and McGrath(2019)]{druckman2019evidence}
Druckman, J.N.; McGrath, M.C.
\newblock The evidence for motivated reasoning in climate change preference
  formation.
\newblock {\em Nat.~Clim.~Change.} {\bf 2019}, {\em 9},~111--119.

\bibitem[Van~der Linden \em{et~al.}(2021)Van~der Linden, Dixon, Clarke, and
  Cook]{van2021inoculating}
Van~der Linden, S.; Dixon, G.; Clarke, C.; Cook, J.
\newblock Inoculating against {COVID-19} vaccine misinformation.
\newblock {\em EClinicalMedicine} {\bf 2021}, {\em 33}.

\bibitem[Horne \em{et~al.}(2015)Horne, Powell, Hummel, and
  Holyoak]{horne2015countering}
Horne, Z.; Powell, D.; Hummel, J.E.; Holyoak, K.J.
\newblock Countering antivaccination attitudes.
\newblock {\em PNAS} {\bf 2015}, {\em 112},~10321--10324.

\bibitem[Nyhan and Reifler(2015)]{nyhan2015does}
Nyhan, B.; Reifler, J.
\newblock Does correcting myths about the flu vaccine work? An experimental
  evaluation of the effects of corrective information.
\newblock {\em Vaccine} {\bf 2015}, {\em 33},~459--464.

\bibitem[Van~Bavel and Pereira(2018)]{van2018partisan}
Van~Bavel, J.J.; Pereira, A.
\newblock The partisan brain: An identity-based model of political belief.
\newblock {\em Trends Cogn. Sci.} {\bf 2018}, {\em 22},~213--224.

\bibitem[Bail \em{et~al.}(2018)Bail, Argyle, Brown, Bumpus, Chen, Hunzaker,
  Lee, Mann, Merhout, and Volfovsky]{Bai18}
Bail, C.A.; Argyle, L.P.; Brown, T.W.; Bumpus, J.P.; Chen, H.; Hunzaker,
  M.B.F.; Lee, J.; Mann, M.; Merhout, F.; Volfovsky, A.
\newblock Exposure to opposing views on social media can increase political
  polarization.
\newblock {\em PNAS} {\bf 2018}, {\em 115},~9216--9221.

\bibitem[Redlawsk(2002)]{redlawsk2002hot}
Redlawsk, D.P.
\newblock Hot cognition or cool consideration? Testing the effects of motivated
  reasoning on political decision making.
\newblock {\em J.~Politics} {\bf 2002}, {\em 64},~1021--1044.

\bibitem[Nyhan and Reifler(2010)]{nyhan2010corrections}
Nyhan, B.; Reifler, J.
\newblock When corrections fail: The persistence of political misperceptions.
\newblock {\em Political Behav.} {\bf 2010}, {\em 32},~303--330.

\bibitem[Nyhan(2021)]{nyhan2021backfire}
Nyhan, B.
\newblock Why the backfire effect does not explain the durability of political
  misperceptions.
\newblock {\em PNAS} {\bf 2021}, {\em 118},~e1912440117.

\bibitem[Chen \em{et~al.}(2021)Chen, Tsaparas, Lijffijt, and
  De~Bie]{chen2021opinion}
Chen, X.; Tsaparas, P.; Lijffijt, J.; De~Bie, T.
\newblock Opinion dynamics with backfire effect and biased assimilation.
\newblock {\em PloS ONE} {\bf 2021}, {\em 16},~e0256922.

\bibitem[Thomm \em{et~al.}(2021)Thomm, Gold, Betsch, and
  Bauer]{thomm2021preservice}
Thomm, E.; Gold, B.; Betsch, T.; Bauer, J.
\newblock When preservice teachers’ prior beliefs contradict evidence from
  educational research.
\newblock {\em Br.~J.~Educ.~Psychol.} {\bf 2021}, {\em 91},~1055--1072.

\bibitem[Cook \em{et~al.}(2004)Cook, Arndt, and Lieberman]{cook2004firing}
Cook, A.; Arndt, J.; Lieberman, J.D.
\newblock Firing back at the backfire effect: The influence of mortality
  salience and nullification beliefs on reactions to inadmissible evidence.
\newblock {\em Law Hum. Behav.} {\bf 2004}, {\em 28},~389--410.

\bibitem[Liebertz and Bunch(2021)]{liebertz2021backfiring}
Liebertz, S.; Bunch, J.
\newblock Backfiring frames: abortion politics, religion, and attitude
  resistance.
\newblock {\em Politics Relig.} {\bf 2021}, {\em 14},~403--430.

\bibitem[Hamilton \em{et~al.}(2015)Hamilton, Hartter, and
  Saito]{hamilton2015trust}
Hamilton, L.C.; Hartter, J.; Saito, K.
\newblock Trust in scientists on climate change and vaccines.
\newblock {\em Sage Open} {\bf 2015}, {\em 5},~2158244015602752.

\bibitem[Kinder and Winter(2001)]{kinder2001exploring}
Kinder, D.R.; Winter, N.
\newblock Exploring the racial divide: Blacks, whites, and opinion on national
  policy.
\newblock {\em Am.~J.~Pol.~Sci.} {\bf 2001}, pp. 439--456.

\bibitem[Lawrence(2022)]{lawrence2022politics}
Lawrence, R.G.
\newblock {\em The Politics of Force: Media and the Construction of Police
  Brutality}; Oxford University Press,  2022.

\bibitem[Hanson(2005)]{hanson2005does}
Hanson, G.H.
\newblock {\em Why does immigration divide America?}; Peterson Institute,
  2005.

\bibitem[Maksymov(2024)]{Mak24_gender}
Maksymov, I.S.
\newblock Magnetism-inspired quantum-mechanical model of gender hluidity.
\newblock {\em Psychol.~J.~Res.~Open.} {\bf 2024}, {\em 6},~1--7.

\bibitem[Tatalovich \em{et~al.}(2014)Tatalovich, Daynes, and
  Lowi]{tatalovich2014moral}
Tatalovich, R.; Daynes, B.W.; Lowi, T.J.
\newblock {\em Moral Controversies in American Politics}; Routledge,  2014.

\bibitem[Abedin(2022)]{abedin2022managing}
Abedin, B.
\newblock Managing the tension between opposing effects of explainability of
  artificial intelligence: a contingency theory perspective.
\newblock {\em Internet Res.} {\bf 2022}, {\em 32},~425--453.

\bibitem[Urban and Pfenning(2000)]{urban2000attitudes}
Urban, D.; Pfenning, U.
\newblock Attitudes towards genetic engineering between change and stability:
  results of a panel study.
\newblock {\em New Genet. Soc.} {\bf 2000}, {\em 19},~251--268.

\bibitem[Montalvo and Reynal-Querol(2003)]{montalvo2003religious}
Montalvo, J.G.; Reynal-Querol, M.
\newblock Religious polarization and economic development.
\newblock {\em Econ.~Lett.} {\bf 2003}, {\em 80},~201--210.

\bibitem[Vicario \em{et~al.}(2019)Vicario, Quattrociocchi, Scala, and
  Zollo]{vicario2019polarization}
Vicario, M.D.; Quattrociocchi, W.; Scala, A.; Zollo, F.
\newblock Polarization and fake news: Early warning of potential misinformation
  targets.
\newblock {\em {ACM} Trans. Web.} {\bf 2019}, {\em 13},~1--22.

\bibitem[Oswald and Grosjean(2004)]{oswald2004confirmation}
Oswald, M.E.; Grosjean, S.
\newblock {\em Confirmation Bias}; Psychology Press: New York, USA,  2004.

\bibitem[Nyhan \em{et~al.}(2013)Nyhan, Reifler, and Ubel]{nyhan2013hazards}
Nyhan, B.; Reifler, J.; Ubel, P.A.
\newblock The hazards of correcting myths about health care reform.
\newblock {\em Medical care} {\bf 2013}, pp. 127--132.

\bibitem[Nyhan \em{et~al.}(2014)Nyhan, Reifler, Richey, and
  Freed]{nyhan2014effective}
Nyhan, B.; Reifler, J.; Richey, S.; Freed, G.L.
\newblock Effective messages in vaccine promotion: a randomized trial.
\newblock {\em Pediatrics} {\bf 2014}, {\em 133},~e835--e842.

\bibitem[Cameron \em{et~al.}(2013)Cameron, Roloff, Friesema, Brown, Jovanovic,
  Hauber, and Baker]{cameron2013patient}
Cameron, K.A.; Roloff, M.E.; Friesema, E.M.; Brown, T.; Jovanovic, B.D.;
  Hauber, S.; Baker, D.W.
\newblock Patient knowledge and recall of health information following exposure
  to ``facts and myths'' message format variations.
\newblock {\em Patient Educ.~Couns.} {\bf 2013}, {\em 92},~381--387.

\bibitem[Thorson(2016)]{thorson2016belief}
Thorson, E.
\newblock Belief echoes: The persistent effects of corrected misinformation.
\newblock {\em Polit.~Commun.} {\bf 2016}, {\em 33},~460--480.

\bibitem[Chan \em{et~al.}(2017)Chan, Jones, Hall~Jamieson, and
  Albarrac{\'\i}n]{chan2017debunking}
Chan, M.p.S.; Jones, C.R.; Hall~Jamieson, K.; Albarrac{\'\i}n, D.
\newblock Debunking: A meta-analysis of the psychological efficacy of messages
  countering misinformation.
\newblock {\em Psychol.~Sci.} {\bf 2017}, {\em 28},~1531--1546.

\bibitem[Swire-Thompson \em{et~al.}(2020)Swire-Thompson, DeGutis, and
  Lazer]{swire2020searching}
Swire-Thompson, B.; DeGutis, J.; Lazer, D.
\newblock Searching for the backfire effect: Measurement and design
  considerations.
\newblock {\em J.~Appl.~Res.~Mem.~Cogn.} {\bf 2020}, {\em 9},~286--299.

\bibitem[Porter \em{et~al.}(2023)Porter, Velez, and Wood]{porter2023correcting}
Porter, E.; Velez, Y.; Wood, T.J.
\newblock Correcting COVID-19 vaccine misinformation in 10 countries.
\newblock {\em R.~Soc.~Open Sci.} {\bf 2023}, {\em 10},~221097.

\bibitem[Ecker \em{et~al.}(2022)Ecker, Lewandowsky, Cook, Schmid, Fazio,
  Brashier, Kendeou, Vraga, and Amazeen]{ecker2022psychological}
Ecker, U.K.; Lewandowsky, S.; Cook, J.; Schmid, P.; Fazio, L.K.; Brashier, N.;
  Kendeou, P.; Vraga, E.K.; Amazeen, M.A.
\newblock The psychological drivers of misinformation belief and its resistance
  to correction.
\newblock {\em Nat.~Rev.~Psychol.} {\bf 2022}, {\em 1},~13--29.

\bibitem[Messiah(1962)]{Mes62}
Messiah, A.
\newblock {\em Quantum Mechanics}; North-Holland Publishing Company, Amsterdam,
   1962.

\bibitem[Busemeyer and Bruza(2012)]{Bus12}
Busemeyer, J.R.; Bruza, P.D.
\newblock {\em Quantum Models of Cognition and Decision}; Oxford University
  Press, New York,  2012.

\bibitem[Khrennikov(2006)]{Khr06}
Khrennikov, A.
\newblock Quantum-like brain: ``Interference of minds''.
\newblock {\em Biosystems} {\bf 2006}, {\em 84},~225--241.

\bibitem[Mindell(2012)]{mindell2012quantum}
Mindell, A.
\newblock {\em Quantum mind: The edge between physics and psychology}; Deep
  Democracy Exchange,  2012.

\bibitem[Khrennikova(2014)]{Khr14}
Khrennikova, P.
\newblock A Quantum Framework for `{Sour Grapes}' in Cognitive Dissonance.
\newblock In Proceedings of the Quantum Interaction; Atmanspacher, H.; Haven,
  E.; Kitto, K.; Raine, D., Eds.; Springer Berlin Heidelberg: Berlin,
  Heidelberg,  2014; pp. 270--280.

\bibitem[Allahverdyan and Galstyan(2014)]{All14}
Allahverdyan, A.E.; Galstyan, A.
\newblock Opinion dynamics with confirmation bias.
\newblock {\em PLoS One} {\bf 2014}, {\em 9},~e99557.

\bibitem[Gronchi and Strambini(2017)]{Gro17}
Gronchi, G.; Strambini, E.
\newblock Quantum cognition and {Bell}’s inequality: A model for
  probabilistic judgment bias.
\newblock {\em J.~Math.~Psychol.} {\bf 2017}, {\em 78},~65--75.

\bibitem[Zhang \em{et~al.}(2017)Zhang, Zhou, Wang, Liu, Shadbolt, Zhang, Gao,
  Li, and O'Brien]{Zha17_1}
Zhang, P.; Zhou, X.Q.; Wang, Y.L.; Liu, B.H.; Shadbolt, P.; Zhang, Y.S.; Gao,
  H.; Li, F.L.; O'Brien, J.L.
\newblock Quantum gambling based on {Nash}-equilibrium.
\newblock {\em Npj Quantum Inf.} {\bf 2017}, {\em 3},~24.

\bibitem[Chen and Hogg(2006)]{Che06}
Chen, K.Y.; Hogg, T.
\newblock How well do people play a quantum {Prisoner's Dilemma?}
\newblock {\em Quantum Inf. Process.} {\bf 2006}, {\em 5},~43--67.

\bibitem[Galam(2005)]{Gal05}
Galam, S.
\newblock Heterogeneous beliefs, segregation, and extremism in the making of
  public opinions.
\newblock {\em Phys.~Rev.~E} {\bf 2005}, {\em 71},~046123.

\bibitem[Castellano \em{et~al.}(2009)Castellano, Fortunato, and Loreto]{Cas09}
Castellano, C.; Fortunato, S.; Loreto, V.
\newblock Statistical physics of social dynamics.
\newblock {\em Rev.~Mod.~Phys.} {\bf 2009}, {\em 81},~591--646.

\bibitem[Hu(2017)]{Hu17}
Hu, H.
\newblock Competing opinion diffusion on social networks.
\newblock {\em R.~Soc.~Open Sci.} {\bf 2017}, {\em 4},~171160.

\bibitem[Eyre \em{et~al.}(2023)Eyre, House, Hill, and Griffiths]{Eyr17}
Eyre, R.W.; House, T.; Hill, E.M.; Griffiths, F.E.
\newblock Spreading of components of mood in adolescent social networks.
\newblock {\em R.~Soc.~Open Sci.} {\bf 2023}, {\em 4},~170336.

\bibitem[Vicario \em{et~al.}(2017)Vicario, Scala, Caldarelli, Stanley, and
  Quattrociocchi]{Del17}
Vicario, M.D.; Scala, A.; Caldarelli, G.; Stanley, H.E.; Quattrociocchi, W.
\newblock Modeling confirmation bias and polarization.
\newblock {\em Sci.~Rep.} {\bf 2017}, {\em 7},~40391.

\bibitem[Redner(2019)]{Red19}
Redner, S.
\newblock Reality-inspired voter models: A mini-review.
\newblock {\em C.~R.~Phys.} {\bf 2019}, {\em 20},~275--292.

\bibitem[Belcastro \em{et~al.}(2020)Belcastro, Cantini, Marozzo, Talia, and
  Trunfio]{Bel20}
Belcastro, L.; Cantini, R.; Marozzo, F.; Talia, D.; Trunfio, P.
\newblock Learning political polarization on social media using neural
  networks.
\newblock {\em IEEE Access} {\bf 2020}, {\em 8},~47177--47187.
\newblock {\url{https://doi.org/10.1109/ACCESS.2020.2978950}}.

\bibitem[Tokita \em{et~al.}(2021)Tokita, Guess, and Tarnita]{Tok21}
Tokita, C.K.; Guess, A.M.; Tarnita, C.E.
\newblock Polarized information ecosystems can reorganize social networks via
  information cascadesi.
\newblock {\em PNAS} {\bf 2021}, {\em 118},~e2102147118.

\bibitem[Cinelli \em{et~al.}(2021)Cinelli, Morales, Galeazzi, Quattrociocchi,
  and Starnini]{Cin21}
Cinelli, M.; Morales, G.D.F.; Galeazzi, A.; Quattrociocchi, W.; Starnini, M.
\newblock The echo chamber effect on social media.
\newblock {\em PNAS} {\bf 2021}, {\em 118},~e2023301118.

\bibitem[Galam and Brooks(2022)]{Gal22}
Galam, S.; Brooks, R.R.W.
\newblock Radicalism: The asymmetric stances of radicals versus conventionals.
\newblock {\em Phys.~Rev.~E} {\bf 2022}, {\em 105},~044112.

\bibitem[Hohmann \em{et~al.}(2023)Hohmann, Devriendt, and Coscia]{Hoh23}
Hohmann, M.; Devriendt, K.; Coscia, M.
\newblock Quantifying ideological polarization on a network using generalized
  {Euclidean} distance.
\newblock {\em Sci.~Adv.} {\bf 2023}, {\em 9},~eabq2044.

\bibitem[Interian and Rodrigues(2023)]{Int23}
Interian, R.; Rodrigues, F.A.
\newblock Group polarization, influence, and domination in online interaction
  networks: a case study of the 2022 {Brazilian} elections.
\newblock {\em J.~Phys.~Complex.} {\bf 2023}, {\em 4},~035008.

\bibitem[Kernell and Lamberson(2023)]{Geo23}
Kernell, G.; Lamberson, P.J.
\newblock Social networks and voter turnout.
\newblock {\em R.~Soc.~Open Sci.} {\bf 2023}, {\em 10},~230704.

\bibitem[Capraro and Perc(2021)]{Cap21}
Capraro, V.; Perc, M.
\newblock Mathematical foundations of moral preferences.
\newblock {\em J.~R.~Soc.~Interface.} {\bf 2021}, {\em 18},~20200880.

\bibitem[Chen \em{et~al.}(2021)Chen, Tsaparas, Lijffijt, and Bie]{Che21}
Chen, X.; Tsaparas, P.; Lijffijt, J.; Bie, T.D.
\newblock Opinion dynamics with backfire effect and biased assimilation.
\newblock {\em PLoS ONE} {\bf 2021}, {\em 16},~e0256922.

\bibitem[Axelrod \em{et~al.}(2021)Axelrod, Daymude, and Forrest]{Axe21}
Axelrod, R.; Daymude, J.J.; Forrest, S.
\newblock Preventing extreme polarization of political attitudes.
\newblock {\em PNAS} {\bf 2021}, {\em 118},~e2102139118.

\bibitem[Aerts and Argu{\"e}lles(2022)]{Aer22}
Aerts, D.; Argu{\"e}lles, J.A.
\newblock Human perception as a phenomenon of quantization.
\newblock {\em Entropy} {\bf 2022}, {\em 24},~1207.

\bibitem[Aerts and Beltran(2022)]{Aer22_1}
Aerts, D.; Beltran, L.
\newblock A {Planck} radiation and quantization scheme for human cognition and
  language.
\newblock {\em Front.~Psychol.} {\bf 2022}, {\em 13}.
\newblock {\url{https://doi.org/10.3389/fpsyg.2022.850725}}.

\bibitem[Kittel(1971)]{Kittel}
Kittel, C.
\newblock {\em Introduction to Solid State Physics}; John Wiley and Sons, New
  York,  1971.

\bibitem[Axelrod(1997)]{Axe97}
Axelrod, R.
\newblock The dissemination of culture: A model with local convergence and
  global polarization.
\newblock {\em J. Conflict Resolut.} {\bf 1997}, {\em 41},~203--226.

\bibitem[Sznajd-Weron and Sznajd()]{Szn00}
Sznajd-Weron, K.; Sznajd, J.
\newblock Opinion evolution in closed community.
\newblock {\em Int.~J.~Mod.~Phys.~C}, {\em 11},~1157--1165.

\bibitem[Rouder and Morey(2009)]{Rou09}
Rouder, J.N.; Morey, R.D.
\newblock The Nature of Psychological Thresholds.
\newblock {\em Psychol.~Rev.} {\bf 2009}, {\em 116},~655--660.

\bibitem[Khrennikov \em{et~al.}(2018)Khrennikov, Basieva, Pothos, and
  Yamato]{Khr18}
Khrennikov, A.; Basieva, I.; Pothos, E.M.; Yamato, I.
\newblock Quantum probability in decision making from quantum information
  representation of neuronal states.
\newblock {\em Sci.~Rep.} {\bf 2018}, {\em 8},~16225.

\bibitem[Lin and Thornton(2023)]{Lin23}
Lin, C.; Thornton, M.
\newblock Evidence for bidirectional causation between trait and mental state
  inferences.
\newblock {\em J.~Exp.~Soc.~Psychol.} {\bf 2023}, {\em 108},~104495.

\bibitem[Toyabe \em{et~al.}(2010)Toyabe, Sagawa, Ueda, Muneyuki, and
  Sano]{Toy10}
Toyabe, S.; Sagawa, T.; Ueda, M.; Muneyuki, E.; Sano, M.
\newblock Experimental demonstration of information-to-energy conversion and
  validation of the generalized {Jarzynski} equality.
\newblock {\em Nat.~Phys.} {\bf 2010}, {\em 6},~988--992.

\bibitem[Dittrich(2014)]{Dit14}
Dittrich, T.
\newblock `The concept of information in physics': an interdisciplinary topical
  lecture.
\newblock {\em Eur.~J.~Phys.} {\bf 2014}, {\em 36},~015010.

\bibitem[Vopson(2019)]{Vop19}
Vopson, M.M.
\newblock The mass-energy-information equivalence principle.
\newblock {\em AIP Adv.} {\bf 2019}, {\em 9},~095206.

\bibitem[Us{\'o}-Dom{\'e}nech and Nescolarde-Selva(2016)]{Uso16}
Us{\'o}-Dom{\'e}nech, J.L.; Nescolarde-Selva, J.
\newblock What are belief systems?
\newblock {\em Found.~Sci.} {\bf 2016}, {\em 21},~147--152.

\bibitem[DeGroot(1974)]{DeG74}
DeGroot, M.H.
\newblock Reaching a consensus.
\newblock {\em J.~Am.~Stat.~Assoc.} {\bf 1974}, {\em 69},~118--121.

\bibitem[Ortega~y Gasset(1966)]{Ortega_Gasset}
Ortega~y Gasset, J.
\newblock {\em Obras Completas, Vol.~I}; Revista de Occidente, Madrid,  1966.

\bibitem[Castro(2013)]{deC13}
Castro, A.D.
\newblock On the quantum principles of cognitive learning.
\newblock {\em Behav. Brain Sci.} {\bf 2013}, {\em 36},~281--282.

\bibitem[Maksymov(2024)]{Mak24_illusions}
Maksymov, I.S.
\newblock Quantum-inspired neural network model of optical illusions.
\newblock {\em Algorithms} {\bf 2024}, {\em 17},~30.

\bibitem[Einstein(1912)]{Ein12}
Einstein, A.
\newblock {Relativit{\"a}t und Gravitation. Erwiderung auf eine Bemerkung von
  M.~Abraham}.
\newblock {\em Annalen der Physik} {\bf 1912}, {\em 343},~1059--1064.

\bibitem[Kragh(1979)]{Kra79}
Kragh, H.
\newblock {{Niels Bohr's} second atomic theory}.
\newblock {\em Hist.~Stud.~Phys.~Sci.} {\bf 1979}, {\em 10},~123--186.

\bibitem[Kuipers(2016)]{Kui16}
Kuipers, T.A.F.
\newblock Models, postulates, and generalized nomic truth approximation.
\newblock {\em Synthese} {\bf 2016}, {\em 193},~3057--3077.

\bibitem[Guerra \em{et~al.}(2021)Guerra, Meira~Jr., Cardie, and
  Kleinberg]{Gue21}
Guerra, P.; Meira~Jr., W.; Cardie, C.; Kleinberg, R.
\newblock A Measure of Polarization on Social Media Networks Based on Community
  Boundaries.
\newblock {\em Proceedings of the International AAAI Conference on Web and
  Social Media} {\bf 2021}, {\em 7},~215--224.
\newblock {\url{https://doi.org/10.1609/icwsm.v7i1.14421}}.

\bibitem[Rollwage \em{et~al.}(2020)Rollwage, Loosen, Hauser, Moran, Dolan, and
  Fleming]{Rol20}
Rollwage, M.; Loosen, A.; Hauser, T.U.; Moran, R.; Dolan, R.J.; Fleming, S.M.
\newblock Confidence drives a neural confirmation bias.
\newblock {\em Nat.~Commun.} {\bf 2020}, {\em 11},~2634.

\bibitem[Moritz \em{et~al.}(2019)Moritz, Klein, Lysaker, and Mehl]{Mor19}
Moritz, S.; Klein, J.P.; Lysaker, P.H.; Mehl, S.
\newblock Metacognitive and cognitive-behavioral interventions for psychosis:
  new developments.
\newblock {\em Dialogues Clin. Neurosci.} {\bf 2019}, {\em 21},~309--317.

\bibitem[Adriaans and van Benthem(2008)]{Adr08}
Adriaans, P.; van Benthem, J.
\newblock {\em Handbook of Philosophy of Information}; Elsevier, Amsterdam,
  2008.

\bibitem[Goss \em{et~al.}(2022)Goss, Morvan, Marinelli, Mitchell, Nguyen, Naik,
  Chen, J{\"u}nger, Kreikebaum, Santiago, Wallman, and Siddiqi]{Gos22}
Goss, N.; Morvan, A.; Marinelli, B.; Mitchell, B.K.; Nguyen, L.B.; Naik, R.K.;
  Chen, L.; J{\"u}nger, C.; Kreikebaum, J.M.; Santiago, D.I.;  et~al.
\newblock High-fidelity qutrit entangling gates for superconducting circuits.
\newblock {\em Nat.~Commun.} {\bf 2022}, {\em 13},~7481.

\bibitem[Halpern \em{et~al.}(2022)Halpern, Ge, and Glendening]{Hal22}
Halpern, A.M.; Ge, Y.; Glendening, E.D.
\newblock Visualizing solutions of the one-dimensional {Schr{\"o}dinger}
  equation using a finite difference method.
\newblock {\em J.~Chem.~Educ.} {\bf 2022}, {\em 99},~3053--3060.

\bibitem[Ortega and Braun(2013)]{Ort13}
Ortega, P.A.; Braun, D.A.
\newblock Thermodynamics as a theory of decision-making with
  information-processing costs.
\newblock {\em Proc.~R.~Soc.~A} {\bf 2013}, {\em 469},~20120683.

\bibitem[Pakhomov and Sudin(2013)]{Pak13}
Pakhomov, A.; Sudin, N.
\newblock Thermodynamic view on decision-making process: emotions as a
  potential power vector of realization of the choice.
\newblock {\em Cogn.~Neurodyn.} {\bf 2013}, {\em 7},~449–463.

\bibitem[Rose-Redwood \em{et~al.}(2018)Rose-Redwood, Kitchin, Rickards, Rossi,
  Datta, and Crampton]{Ros18}
Rose-Redwood, R.; Kitchin, R.; Rickards, L.; Rossi, U.; Datta, A.; Crampton, J.
\newblock The possibilities and limits to dialogue.
\newblock {\em Dialogues Hum.~Geogr.} {\bf 2018}, {\em 8},~109--123.

\bibitem[Putnam(2022)]{Put22}
Putnam, H.
\newblock {\em Philosophy as Dialogue}; Harvard University Press: Cambridge, MA
  and London, England,  2022.
\newblock {\url{https://doi.org/doi:10.4159/9780674287594}}.

\bibitem[Kovbasyuk(2011)]{Kov11}
Kovbasyuk, O.
\newblock {Dialogue as a means of change}.
\newblock {\em Intl.~HETL Rev.} {\bf 2011}, {\em 1},~2.

\bibitem[Larcinese \em{et~al.}(2013)Larcinese, Snyder, and Testa]{Lar13}
Larcinese, V.; Snyder, J.M.; Testa, C.
\newblock Testing models of distributive politics using exit polls to measure
  voters' preferences and partisanship.
\newblock {\em Br.~J.~Political Sci.} {\bf 2013}, {\em 43},~845--875.

\bibitem[Jovanovic and Leeuwen(2018)]{Jov18}
Jovanovic, D.; Leeuwen, T.V.
\newblock Multimodal dialogue on social media.
\newblock {\em Soc.~Semiot.} {\bf 2018}, {\em 28},~683--699.

\bibitem[Ricaud \em{et~al.}(2019)Ricaud, Borgnat, Tremblay, Gon{\c{c}}alves,
  and Vandergheynst]{Ric19}
Ricaud, B.; Borgnat, P.; Tremblay, N.; Gon{\c{c}}alves, P.; Vandergheynst, P.
\newblock Fourier could be a data scientist: From graph {Fourier} transform to
  signal processing on graphs.
\newblock {\em C.~R.~Phys.} {\bf 2019}, {\em 20},~474--488.

\bibitem[Baszuro and Swacha(2021)]{Bas21}
Baszuro, P.; Swacha, J.
\newblock Graph analysis using fast {Fourier} transform applied on grayscale
  bitmap images.
\newblock {\em Information} {\bf 2021}, {\em 12},~454.

\bibitem[Mcculloh \em{et~al.}(2012)Mcculloh, Johnson, and Carley]{Mcc12}
Mcculloh, I.A.; Johnson, A.N.; Carley, K.M.
\newblock Spectral analysis of social networks to identify periodicity.
\newblock {\em J. Math. Sociol.} {\bf 2012}, {\em 36},~80--96.

\bibitem[Vasudevan \em{et~al.}(2015)Vasudevan, Belianinov, Gianfrancesco,
  Baddorf, Tselev, Kalinin, and Jesse]{Vas15}
Vasudevan, R.K.; Belianinov, A.; Gianfrancesco, A.G.; Baddorf, A.P.; Tselev,
  A.; Kalinin, S.V.; Jesse, S.
\newblock {Big data in reciprocal space: Sliding fast {Fourier} transforms for
  determining periodicity}.
\newblock {\em Appl.~Phys.~Lett.} {\bf 2015}, {\em 106},~091601.

\bibitem[Parrillo and Donoghue(2005)]{Par19}
Parrillo, V.N.; Donoghue, C.
\newblock Updating the {Bogardus} social distance studies: a new national
  survey.
\newblock {\em Soc.~Sci.~J.} {\bf 2005}, {\em 42},~257--271.

\bibitem[Antonopoulos and Shang(2018)]{Ant18}
Antonopoulos, C.G.; Shang, Y.
\newblock Opinion formation in multiplex networks with general initial
  distributions.
\newblock {\em Sci.~Rep.} {\bf 2018}, {\em 8},~2852.

\bibitem[Kuikka(2023)]{Kui23}
Kuikka, V.
\newblock Opinion Formation on social networks---The effects of recurrent and
  circular influence.
\newblock {\em Computation} {\bf 2023}, {\em 11},~103.

\bibitem[Nugent \em{et~al.}(2023)Nugent, Gomes, and Wolfram]{Nug23}
Nugent, A.; Gomes, S.N.; Wolfram, M.T.
\newblock On evolving network models and their influence on opinion formation.
\newblock {\em Phys.~D} {\bf 2023}, {\em 456},~133914.

\bibitem[Vannucci and Ohannessian(2019)]{Van19}
Vannucci, A.; Ohannessian, C.M.
\newblock {Social media use bubgroups differentially predict psychosocial
  well-being during early adolescence}.
\newblock {\em J.~Youth Adolesc.} {\bf 2019}, {\em 48},~1469--1493.

\bibitem[Imbens and Rubin(2015)]{Imb15}
Imbens, G.; Rubin, D.
\newblock {\em Causal Inference for Statistics, Social, and Biomedical
  Sciences: An Introduction}; Cambridge University Press,  2015.

\bibitem[Jusup \em{et~al.}(2022)Jusup, Holme, Kanazawa, Takayasu, Romi{\'c},
  Wang, Ge{\v{c}}ek, Lipi{\'c}, Podobnik, Wang, Luo, Klanj{\v{s}}{\v{c}}ek,
  Fan, Boccaletti, and Perc]{Jus22}
Jusup, M.; Holme, P.; Kanazawa, K.; Takayasu, M.; Romi{\'c}, I.; Wang, Z.;
  Ge{\v{c}}ek, S.; Lipi{\'c}, T.; Podobnik, B.; Wang, L.;  et~al.
\newblock Social physics.
\newblock {\em Phys.~Rep.} {\bf 2022}, {\em 948},~1--148.

\bibitem[Galam(2022)]{Gal22_1}
Galam, S.
\newblock Physicists, non physical topics, and interdisciplinarity.
\newblock {\em Front.~Phys.} {\bf 2022}, {\em 10}.
\newblock {\url{https://doi.org/10.3389/fphy.2022.986782}}.

\bibitem[Galam(2017)]{Gal17}
Galam, S.
\newblock The {Trump} phenomenon: An explanation from sociophysics.
\newblock {\em Int.~J.~Mod.~Phys.~B} {\bf 2017}, {\em 31},~1742015.

\bibitem[Galam(2018)]{Gal18}
Galam, S.
\newblock Unavowed abstention can overturn poll predictions.
\newblock {\em Front.~Phys.} {\bf 2018}, {\em 7}.
\newblock {\url{https://doi.org/10.3389/fphy.2018.00024}}.

\bibitem[Nielsen and Chuang(2002)]{Nie02}
Nielsen, M.; Chuang, I.
\newblock {\em Quantum Computation and Quantum Information}; Oxford University
  Press, New York,  2002.

\bibitem[Pothos and Busemeyer(2009)]{Pot09}
Pothos, E.M.; Busemeyer, J.R.
\newblock A quantum probability explanation for violations of `rational'
  decision theory.
\newblock {\em Proc.~R.~Soc.~B} {\bf 2009}, {\em 276},~2171--2178.

\bibitem[Benedek and Caglioti(2019)]{Ben18}
Benedek, G.; Caglioti, G.
\newblock Graphics and Quantum Mechanics--The Necker Cube as a Quantum-like
  Two-Level System.
\newblock In Proceedings of the Proceedings of the 18th International
  Conference on Geometry and Graphics; Cocchiarella, L., Ed. Springer
  International Publishing,  2019, pp. 161--172.

\bibitem[Pothos and Busemeyer(2022)]{Pot22}
Pothos, E.M.; Busemeyer, J.R.
\newblock Quantum Cognition.
\newblock {\em Annu.~Rev.~Psychol.} {\bf 2022}, {\em 73},~749--778.

\bibitem[Atmanspacher and Filk(2010)]{Atm10}
Atmanspacher, H.; Filk, T.
\newblock A proposed test of temporal nonlocality in bistable perception.
\newblock {\em J.~Math.~Psychol.} {\bf 2010}, {\em 54},~314--321.

\bibitem[Baaquie(2020)]{Baa20}
Baaquie, B.E.
\newblock {\em Mathematical Methods and Quantum Mathematics for Economics and
  Finance}; Springer, Singapore,  2020.

\bibitem[Herman \em{et~al.}(2023)Herman, Googin, Liu, Sun, Galda, Safro,
  Pistoia, and Alexeev]{Her23}
Herman, D.; Googin, C.; Liu, X.; Sun, Y.; Galda, A.; Safro, I.; Pistoia, M.;
  Alexeev, Y.
\newblock Quantum computing for finance.
\newblock {\em Nat.~Rev.~Phys.} {\bf 2023}, {\em 5},~450--465.

\bibitem[Northcott(2009)]{Nor09}
Northcott, A.
\newblock {\em The Complete Guide to Using Candlestick Charting: How to Earn
  High Rates of Return---Safely}; Atlantic Publishing Group,  2009.

\bibitem[Pedersen(2022)]{Ped22}
Pedersen, L.H.
\newblock Game on: Social networks and markets.
\newblock {\em J.~Financ.~Econ.} {\bf 2022}, {\em 146},~1097--1119.

\bibitem[Hirshleifer \em{et~al.}(2023)Hirshleifer, Peng, and Wang]{Hir23}
Hirshleifer, D.; Peng, L.; Wang, Q.
\newblock News diffusion in social networks and stock market reactions.
\newblock {\em Working paper of National Bureau of Ecomonic Research} {\bf
  2023}, {\em 30860}.
\newblock {\url{https://doi.org/10.3386/w30860}}.

\bibitem[Fabretti(2022)]{Fab22}
Fabretti, A.
\newblock A dynamical model for financial market: among common market
  strategies who and how moves the price to fluctuate, inflate, and burst?
\newblock {\em Mathematics} {\bf 2022}, {\em 10}.

\bibitem[McKenna \em{et~al.}(1994)McKenna, McMullen, and Shlesinger]{Mck94}
McKenna, T.M.; McMullen, T.A.; Shlesinger, M.F.
\newblock The brain as a dynamic physical system.
\newblock {\em Neuroscience} {\bf 1994}, {\em 60},~587--605.

\bibitem[Korn and Faure(2003)]{Kor03}
Korn, H.; Faure, P.
\newblock Is there chaos in the brain? II. Experimental evidence and related
  models.
\newblock {\em C.~R.~Biol.} {\bf 2003}, {\em 326},~787--840.

\bibitem[Garcia \em{et~al.}(2020)Garcia, {Machado Pereira}, Acebal, and {Bosco
  de Magalh{\~a}es}]{Gar20}
Garcia, M.M.; {Machado Pereira}, A.C.; Acebal, J.L.; {Bosco de Magalh{\~a}es},
  A.R.
\newblock Forecast model for financial time series: An approach based on
  harmonic oscillators.
\newblock {\em Phys.~A} {\bf 2020}, {\em 549},~124365.

\bibitem[Ahn \em{et~al.}(2018)Ahn, Choi, Dai, Sohn, and Yang]{Ahn18}
Ahn, K.; Choi, M.Y.; Dai, B.; Sohn, S.; Yang, B.
\newblock Modeling stock return distributions with a quantum harmonic
  oscillator.
\newblock {\em Europhys.~Lett.~(EPL)} {\bf 2018}, {\em 120},~38003.

\bibitem[Shiratori \em{et~al.}(2023)Shiratori, Shinkawa, R{\"o}hm, Chauvet,
  Segawa, Laurent, Bachelier, Yamagami, Horisaki, and Naruse]{Shi23}
Shiratori, H.; Shinkawa, H.; R{\"o}hm, A.; Chauvet, N.; Segawa, E.; Laurent,
  J.; Bachelier, G.; Yamagami, T.; Horisaki, R.; Naruse, M.
\newblock Asymmetric quantum decision-making.
\newblock {\em Sci.~Rep.} {\bf 2023}, {\em 13},~14636.

\bibitem[Maksymov and Pogrebna(2023)]{Mak24_preference}
Maksymov, I.S.; Pogrebna, G.
\newblock The Physics of Preference: Unravelling Imprecision of Human
  Preferences through Magnetisation Dynamics,  2023,
  \href{http://xxx.lanl.gov/abs/2310.00267}{{\normalfont
  [arXiv:physics.soc-ph/2310.00267]}}.

\bibitem[Christensen \em{et~al.}(2020)Christensen, Loh, Picek, Jakobovi{\'c},
  Jing, Fisher, Ceperic, Joannopoulos, and Solja{\v{c}}i{\'c}]{Chr20}
Christensen, T.; Loh, C.; Picek, S.; Jakobovi{\'c}, D.; Jing, L.; Fisher, S.;
  Ceperic, V.; Joannopoulos, J.D.; Solja{\v{c}}i{\'c}, M.
\newblock Predictive and generative machine learning models for photonic
  crystals.
\newblock {\em Nanophotonics} {\bf 2020}, {\em 9},~4183--4192.

\bibitem[Orabi \em{et~al.}(2020)Orabi, Mouheb, {Al Aghbari}, and Kamel]{Ora20}
Orabi, M.; Mouheb, D.; {Al Aghbari}, Z.; Kamel, I.
\newblock Detection of bots in cocial media: A systematic review.
\newblock {\em Inf.~Process.~Manage.} {\bf 2020}, {\em 57},~102250.

\bibitem[Hayawi \em{et~al.}(2023)Hayawi, Saha, Masud, Mathew, and
  Kaosar]{Hay23}
Hayawi, K.; Saha, S.; Masud, M.M.; Mathew, S.S.; Kaosar, M.
\newblock Social media bot detection with deep learning methods: a systematic
  review.
\newblock {\em Neural Comput.~Applic.} {\bf 2023}, {\em 35},~8903--8918.

\bibitem[Foster and Efthymiou(2022)]{KPMG}
Foster, C.; Efthymiou, T.
\newblock Operational optimisation: Decision-making beyond human capability.
\newblock {\em KPMG Report} {\bf 2022}, pp. 1--25.

\bibitem[Taddeo(2016)]{Tad16}
Taddeo, M.
\newblock On the risks of relying on analogies to understand cyber conflicts.
\newblock {\em Minds Mach.} {\bf 2016}, {\em 26},~317--321.

\bibitem[Rashkovskiy(2019)]{Ras19}
Rashkovskiy, S.A.
\newblock `Bosons' and `fermions' in social and economic systems.
\newblock {\em Phys.~A} {\bf 2019}, {\em 514},~90--104.

\bibitem[Martinez-Perdiguero(2020)]{Per20}
Martinez-Perdiguero, J.
\newblock Fermion-like behavior of elements/agents in their spatial
  distribution around points of interest.
\newblock {\em Phys.~A} {\bf 2020}, {\em 557},~124905.

\bibitem[Tung(1985)]{Tun85}
Tung, W.K.
\newblock {\em Group Theory in Physics}; World Scientific, Singapore,  1985.

\bibitem[Miyazawa(2010)]{Miy10}
Miyazawa, H.
\newblock Superalgebra and fermion-boson symmetry.
\newblock {\em Proc.~Jpn.~Acad.~Ser.~B Phys.~Biol.~Sci.} {\bf 2010}, {\em
  86},~158--164.

\bibitem[Taie \em{et~al.}(2022)Taie, Ibarra-Garc{\'i}a-Padilla, Nishizawa,
  Takasu, Kuno, Wei, Scalettar, Hazzard, and Takahashi]{Tai22}
Taie, S.; Ibarra-Garc{\'i}a-Padilla, E.; Nishizawa, N.; Takasu, Y.; Kuno, Y.;
  Wei, H.T.; Scalettar, R.T.; Hazzard, K.R.A.; Takahashi, Y.
\newblock Observation of antiferromagnetic correlations in an ultracold
  {SU}($N$) {Hubbard} model.
\newblock {\em Nat.~Phys.} {\bf 2022}, {\em 18},~1356--1361.

\bibitem[Sonderhouse \em{et~al.}(2020)Sonderhouse, Sanner, Hutson, Goban,
  Bilitewski, Yan, Milner, Rey, and Ye]{Son20}
Sonderhouse, L.; Sanner, C.; Hutson, R.B.; Goban, A.; Bilitewski, T.; Yan, L.;
  Milner, W.R.; Rey, A.M.; Ye, J.
\newblock Thermodynamics of a deeply degenerate SU($N$)--symmetric {Fermi} gas.
\newblock {\em Nat.~Phys.} {\bf 2020}, {\em 16},~1216--1221.

\end{thebibliography}

\end{adjustwidth}
\end{document}